%% file: main.tex
\documentclass[11pt]{article}

\usepackage{graphicx}
\usepackage{subfigure}
\usepackage{amsmath}
\usepackage{mathtools}
\usepackage{amssymb}
\usepackage{xfrac}
\usepackage{siunitx}

\input{mymacros}

\title{Role of anisotropic strength and stiffness in governing the initiation and propagation of yielding in polycrystalline solids}
\author{Andrew C. Poshadel and Paul R. Dawson \\
Sibley School of Mechanical and Aerospace Engineering\\
 Cornell University}
\date{}

\begin{document}
\maketitle

\begin{abstract}
The ratio of directional strength-to-stiffness is important in governing the relative order in which individual crystals within a polycrystalline aggregate will yield as the aggregate is loaded.  In this paper, a strength-to-stiffness parameter is formulated for multiaxial loading that extends the development  of Wong and Dawson~\cite{Wong10a} for uniaxial loading. Building on the principle of strength-to-stiffness, a methodology for predicting the macroscopic stresses at which elements in a finite element mesh yield is developed. This analysis uses elastic strain data from one increment of a purely elastic finite element simulation to make the prediction, given knowledge of the single-crystal yield surface. Simulations of austenitic strainless steel AL6XN are used to demonstrated the effectivness of the strength-to-stiffness parameter and yield prediction methodology.    
\end{abstract}
\pagebreak[4]

\section{Introduction}
\label{sec:introduction}
\input{Introduction}

\section{Formulation of the strength-to-stiffness parameter}
\label{sec:y2e_formulation}
\input{Y2E_Formulation}
\section{Constitutive model equations}
\label{sec:model_methods}
\input{Model_Methods}
\section{Material, Virtual Sample and Loading}
\label{sec:virtual_polycrystal}
\input{material_and_virtualpolycrystal}

\section{Strength-to-stiffness analysis of austenitic stainless steel}
\label{sec:y2e_application}
\input{Y2E_Application}

\section{Methodology for predicting the macroscopic stress at which yielding occurs locally at the microscale}
\label{sec:yield_predict_formulation}
\input{Yield_Predict_Formulation}
\section{Application of yield prediction methodology to austenitic stainless steel}
\label{sec:yield_predict_application}
\input{Yield_Predict_Application}

\section{Conclusions}
\label{sec:conclusions}
\input{Conclusions}

\section{Acknowledgements}
\label{sec:acknowledgements}
\input{Acknowledgements}

\bibliographystyle{unsrt}
\bibliography{References}
\appendix
\section{Finite Element Formulation for the Motion}
\label{sec:appendix}
\input{appendix}

\end{document}

%% file: mymacros.tex
\DeclareMathAlphabet{\mathsfsl}{OT1}{cmss}{m}{sl}

\DeclareMathOperator{\trace}{tr}

\newcommand{\fepx}{{\bfseries{\slshape{FEpX}}}}

\newcommand{\vctr}[1]{\boldsymbol{#1}}

\newcommand{\tensSym}[1]{\boldsymbol{#1}}

\newcommand{\tnsr}[1]{\mathsfsl{#1 }}

\newcommand{\gtnsr}[1]{\mathsfsl{#1}}

\newcommand{\rodvec}{\vctr{r}}

\newcommand{\gradop}{{\rm grad }}
\newcommand{\divop}{{\rm div}}

\newcommand{\vectop}{{\rm vect}}
\newcommand{\transp}{{\rm T}}
\newcommand{\invrs}{{ -\hspace{-.1em}1}}
\newcommand{\deeteei }{ \frac{1}{\Delta t}  }   
\newcommand{\betaoverkappa}{ \frac{\beta}{\kappa \Delta t}  }

\newcommand{\cauchy}{{\boldsymbol{\sigma}}}
\newcommand{\dcauchy}{{\boldsymbol{\sigma}}^\prime}

\newcommand{\pxspinhat}{\hat{\tnsr{w}}^{p}}
\newcommand{\spinvec}{\vctr{\omega}}

\newcommand{\skwschmid}{\hat{\tnsr{q}}^\alpha}

\newcommand{\rss}{\tau^\alpha}
\newcommand{\gammadot}{\dot{\gamma}^\alpha}
\newcommand{\gdotz}{\dot{\gamma}_0}

\newcommand{\sumss}{\sum_{\alpha}}
\newcommand{\matlatepsdot}{ \left\{ \dot{ \mathsf{e}^e} \right\} }   
   
\newcommand{\matlateps }{ \left\{\mathsf{e}^e\right\}  }   
\newcommand{\matlatepsold }{ \left\{\mathsf{e}_0^e\right\}  }  
\newcommand{\matdlateps }{ \left\{ {\mathsf{e}^e}^\prime \right\}  }    
\newcommand{\matdlatepsold }{ \left\{ {\mathsf{e}_0^e}^\prime \right\}  }  
\newcommand{\matdefrate }{\Big\{ \mathsf{d} \Big\}   }     
  
\newcommand{\matddefrate }{\Big\{ \mathsf{d}^\prime \Big\}   }
\newcommand{\matlatdefrate }{\Big\{ {\hat{\mathsf{d}}^p} \Big\}   }
\newcommand{\matpxspinhat }{\Big[ {\hat{\mathsf{w}}^p} \Big]   }
  
\newcommand{\matdkirch }{  \left\{ \tau^\prime \right\}  }   
  
\newcommand{\matdcauchy }{  \left\{ \sigma^\prime \right\}  } 
\newcommand{\matxdelasticity}{\Big[ \,\mathsf{c}^\prime \,\Big]}  
  
\newcommand{\matxplasticity}{\Big[ \,\mathsf{m} \,\Big]}  
\newcommand{\matxep}{\Big[ \,\mathsf{s} \,\Big]}

\newcommand{\matsymschmid }{ \Big\{\mathsf{p}^\alpha\Big\}  }  
\newcommand{\mathhh}{\Big\{ \mathsf{h} \Big\}   }   
   
\newcommand{\matbodyforce}{\Big\{ \iota \Big\}   }

\newcommand{\matvel }{ \Big\{ v \Big\}  } 
\newcommand{\matcapX}{\Big[ \,\mathsf{X} \,\Big]}    
\newcommand{\matcapB}{\Big[ \,\mathsf{B} \,\Big]}
\newcommand{\matcapN}{\Big[ \,\mathsf{N}(\xi, \eta, \zeta) \,\Big]}

\newcommand{\matresiduale}{\Big\{ R^{\it ele}_u \Big\}}
\newcommand{\matvelnp}{\Big\{ \mathsf{V} \Big\}}

\newcommand{\matstiffd}{\Big[ \,\mathsf{k}^{\it ele}_d \,\Big]} 
\newcommand{\matstiffv}{\Big[ \,\mathsf{k}^{\it ele}_v \,\Big]}

\newcommand{\matdelta}{\Big\{ \mathsf{\delta} \Big\}}

\newcommand{\dee}{{\mathrm{d}}}

%% file: Introduction.tex
Crystals exhibit both plastic and elastic anisotropy. Plastic deformation due to crystallographic slip occurs on a restricted set of slip systems. For a prescribed stress state, some crystallographic orientations are more favorable for slip than others. Thus, the directional strength of a crystal exhibits orientation dependence. Crystals are also elastically anisotropic, and the elastic response is likewise dependent on crystallographic orientation.

For uniaxial loading, the orientational dependence of strength is typically characterized by either the Schmid or Taylor factor. The Schmid factor is based on an isostress assumption, which satisfies equilibrium, whereas the Taylor factor is based on an isostrain assumption, which satisfies compatibility. Both factors relate the macroscopic yield stress to the critical resolved shear stress on a slip system. The elastic response is characterized by the directional modulus.

Wong and Dawson~\cite{Wong10a} demonstrated that for uniaxial loading, it is the ratio of directional strength to directional stiffness, rather than the pure directional strength, which correlates to the order in which crystals yield. Stiffness is important because of the deformation compatibility constraints imposed on a grain in a polycrystalline aggregate by its neighbors. To illustrate this point, Wong and Dawson presented a simple analog system of two materials loaded in parallel between two rigid plates. In this system, it is precisely the ratio of strength-to-stiffness that  governs which material yields first. A stiff, strong material can yield before a compliant, weak material. 

They developed two formulations for the strength-to-stiffness parameter. The first, which they termed the single-crystal strength-to-stiffness, was formulated for isolated single crystals. The strength-to-stiffness ratio is defined as the ratio of the reciprocal Schmid factor to the directional modulus. This form of strength-to-stiffness can be evaluated analytically can used as a prescriptive measure for when a crystal will yield. However, it does not account for the fact that local stress states are different from the macroscopic stress state as a result of intergranular interactions. A second strength-to-stiffness formulation was developed to include the effects of intergranular interactions. This parameter was evaluated using an elasto-plastic finite element simulation, and is therefore descriptive rather than predictive. They termed this parameter the fiber-averaged strength-to-stiffness, because they averaged the quantity along crystallographic fibers. However, its general formulation does not depend on fiber averaging. It can be evaluated for individual finite elements as well. In this work, it is therefore referred to as the simulated strength-to-stiffness. In this formulation, the Taylor factor is evaluated in the fully-developed plastic regime from simulated data. The Taylor factor is defined as the ratio of the sum of the slip system shear rate magnitudes to the effective deformation rate. The simulated strength-to-stiffness parameter is defined as the ratio of the Taylor factor to the directional modulus. Wong and Dawson demonstrate good correlation between strength-to-stiffness and the order in which crystals yield. The simulated strength-to-stiffness, which incorporates differences in local stress state due to intergranular effects, exhibits better correlation than the single crystal strength-to-stiffness. However, both measures exhibit similar qualitative trends.

In this work, a new strength-to-stiffness parameter is formulated for multiaxial loading. The analysis is similar to that of Wong and Dawson~\cite{Wong10a}, who showed that the ratio of directional strength-to-stiffness is important in governing the relative order in which crystals yield for uniaxial loading. Building on the principle of strength-to-stiffness, a methodology for predicting the macroscopic stresses at which elements in a finite element mesh yield is developed. This analysis uses elastic strain data from one increment of a purely elastic finite element simulation to make the prediction, given knowledge of the single-crystal yield surface. Simulations of austenitic strainless steel AL6XN are used to demonstrated the effectivness of the strength-to-stiffness parameter and yield prediction methodology.

%% file: Y2E_Formulation.tex
The strength-to-stiffness analysis of Wong and Dawson is expanded to include generalized multiaxial loading. The new formulation of strength-to-stiffness ratio utilizes local elastic strains, computed from one load increment of a purely elastic finite element simulation. The use of simulated elastic strains incorporates information about the local stress state in the purely elastic regime. Even though the formulation utilizes simulation data, it is still a predictive measure for the order in which crystals yield. In the present formulation, residual stresses are assumed to be zero. Furthermore, only monotonic loading, in which the principal directions of the applied macroscopic stress remain fixed and the ratios of principal stress components are constant, is considered. 

\begin{figure}[b]
\centering
\includegraphics{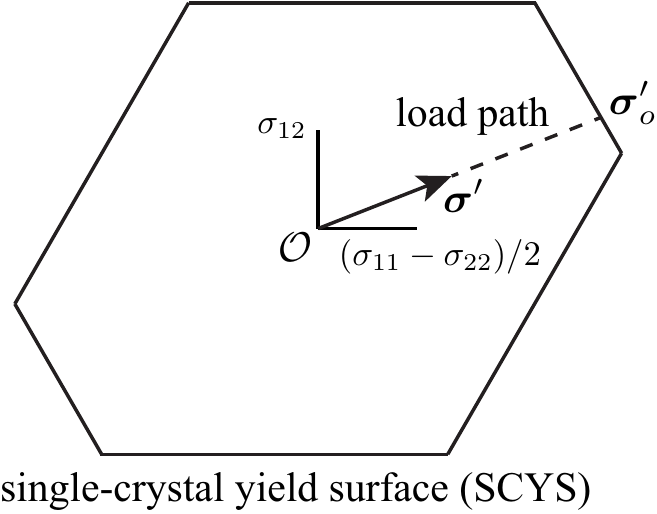}
\caption{Schematic of a two-dimensional single-crystal yield surface (SCYS). In general, deviatoric stress-space is five-dimensional. For a linear stress path, the local yield stress $\tensSym{\sigma}^\prime_o$ is a scalar multiple of the local deviatoric stress $\tensSym{\sigma}^\prime$.}
\label{fig:StressSchematic}
\end{figure}

For this type of loading, the macroscopic stress tensor $\tensSym{\Sigma}$ can be written as the product of a scalar coefficient $\Sigma$ and a basis tensor $\hat{\tensSym{\Sigma}}$
\begin{equation}\label{eqn:y2e_MonotonicLoad}
\tensSym{\Sigma} = \Sigma \hat{\tensSym{\Sigma}}
\end{equation}
The scalar coefficient can be thought of as a stress magnitude and the basis tensor as a stress direction. In this derivation, all basis tensors are denoted with a hat. The basis tensor for the macroscopic stress is constructed such that its largest eigenvalue magnitude is equal to unity. The coefficient $\Sigma$ is positive and increases monotonically with time. Similarly the local stress can also be written as the product of a scalar coefficient $\sigma$ and a basis tensor $\hat{\tensSym{\sigma}}$
\begin{equation}\label{eqn:y2e_LocalStress}
\tensSym{\sigma} = \sigma \hat{\tensSym{\sigma}}
\end{equation}
where $\hat{\tensSym{\sigma}}$ is the local stress corresponding to a macroscopic stress of $\hat{\tensSym{\Sigma}}$. In the purely elastic regime, the local and macroscopic scalar coefficients are equal by construction
\begin{equation}\label{eqn:y2e_CoeffEqual}
\sigma = \Sigma
\end{equation}
 Note that, unlike the macroscopic stress basis tensor, $\hat{\tensSym{\Sigma}}$, largest eigenvalue of the local stress basis tensor, $\hat{\tensSym{\sigma}}$, is not, in general, equal to unity. 
Since the local stress is a spatially-varying field quantity, the basis tensor for the local stress $\hat{\tensSym{\sigma}}$ is also spatially varying. Once yielding begins in an aggregate, Equation~\ref{eqn:y2e_CoeffEqual} no longer holds. However, for the present analysis, it is assumed that the principal directions of the local stress tensor and the ratios of the local principal stress components remain constant, until the local stress is on the single-crystal yield surface (SCYS). This assumption implies a linear path through deviatoric stress space. The local yield stress $\tensSym{\sigma}^\prime_o$ is defined as the point where the stress path intersects the SCYS. The SCYS, stress path, and local yield stress are illustrated schematically for a two-dimensional stress state in Figure~\ref{fig:StressSchematic}. Note that a general deviatoric stress state is five-dimensional.
The von Mises stress defines a norm on the deviatoric stress space and represents the distance from the origin $\mathcal{O}$ to $\tensSym{\sigma}^\prime$. 
\begin{equation}\label{eqn:y2e_vonMises}
\Vert \tensSym{\sigma}^\prime \Vert = \sqrt{\sfrac{3}{2}} \left( \tensSym{\sigma}^\prime : \tensSym{\sigma}^\prime \right)^{\sfrac{1}{2}}
\end{equation}

Yielding corresponds to the onset of plastic deformation due to crystallographic slip. Crystallographic slip occurs on a restricted set of slip systems, where $\alpha$ denotes the slip system index. The resolved shear stress on the $\alpha$-slip system $\tau^\alpha$ is the projection of the local deviatoric stress onto the slip system
\begin{equation}\label{eqn:y2e_RSS}
\tau^\alpha = \tensSym{P}^\alpha:\tensSym{\sigma}^\prime
\end{equation}
where $\tensSym{P}^\alpha$ is the symmetric part of the Schmid tensor. In the rate-independent limit, yielding occurs when the magnitude of the resolved shear stress on any slip system is equal to the critical resolved shear stress $\tau^\alpha_{cr}$ for that slip system. It is assumed that there is no Bauschinger effect and that each slip system is equally resistant to slip in the forward and reverse directions. The yield condition is given by 
\begin{equation}\label{eqn:y2e_YieldCondition}
\max_\alpha \left( \frac{\vert \tau^\alpha \vert}{\tau^\alpha_{cr}} \right) = 1
\end{equation}
Let $(\cdot)^*$ denote a slip system where 
\begin{equation}
\label{eqn:y2e_TauStar}
\frac{\vert \tau^* \vert}{\tau^*_{cr}} \equiv \max_\alpha \left( \frac{\vert \tau^\alpha \vert}{\tau^\alpha_{cr}} \right)
\end{equation}
By construction, the local yield stress $\tensSym{\sigma}^\prime_o$ is a scalar multiple of the local deviatoric stress $\tensSym{\sigma}^\prime$ for a linear stress path. The two stresses are related by a scaling of $\vert \tau^* \vert/ \tau^*_{cr}$
\begin{equation}
\label{eqn:y2e_StressScale}
\tensSym{\sigma}^\prime = \frac{\vert \tau^* \vert}{\tau^*_{cr}} \tensSym{\sigma}^\prime_o
\end{equation}
This can be proven by taking the projection of the local yield stress in Equation~\ref{eqn:y2e_StressScale} onto a slip system corresponding to $\tau^*$ and rearranging terms
\begin{equation}\label{eqn:y2e_ScaleProof2}
\frac{\tensSym{P}^* : \tensSym{\sigma}^\prime}{\tensSym{P}^* : \tensSym{\sigma}^\prime_o} = \frac{\vert \tau^* \vert}{\tau^*_{cr}}
\end{equation}
Since $\tensSym{\sigma}^\prime$ and $\tensSym{\sigma}^\prime_o$ are related by a scaling, the lefthand side of Equation~\ref{eqn:y2e_ScaleProof2} is positive. Therefore,
\begin{equation}
\label{eqn:y2e_ScaleProof3}
\left| \frac{\tensSym{P}^* : \tensSym{\sigma}^\prime}{\tensSym{P}^* : \tensSym{\sigma}^\prime_o} \right| = \frac{\vert \tau^* \vert}{\tau^*_{cr}}
\end{equation}
Evaluating the inner products on the lefthand side produces the equality
\begin{equation}\label{eqn:y2e_ScaleProof4}
\frac{\vert \tau^* \vert}{\tau^*_{cr}} = \frac{\vert \tau^* \vert}{\tau^*_{cr}}
\end{equation}
An interpretation of the ratio $\vert \tau^* \vert/ \tau^*_{cr}$ can be derived by taking the norm of both sides of Equation~\ref{eqn:y2e_StressScale} and rearranging terms
\begin{equation}\label{eqn:y2e_StressRatio}
\frac{\vert \tau^* \vert}{\tau^*_{cr}} = \frac{\Vert \tensSym{\sigma}^\prime \Vert}{\Vert \tensSym{\sigma}^\prime_o \Vert}
\end{equation}
In deviatoric stress space, the ratio $\vert \tau^* \vert/ \tau^*_{cr}$ represents the distance from the origin to the local deviatoric stress $\Vert\tensSym{\sigma}^\prime \Vert$, normalized by the distance from the origin to the SCYS along a linear loading path $\Vert\tensSym{\sigma}^\prime_o \Vert$. It is a measure of proximity to the SCYS along the loading path, with a value of 0 corresponding to being at the origin of the SCYS and a value of 1 corresponding to being on the SCYS. Regions with high $\vert \tau^* \vert/ \tau^*_{cr}$ tend to yield before regions with low $\vert \tau^* \vert/ \tau^*_{cr}$.

The strength-to-stiffness ratio $r_{SE}$, which governs the onset of yielding, is related to $\vert \tau^* \vert/ \tau^*_{cr}$ and has the form
\begin{equation}\label{eqn:y2e_Y2E}
r_{SE}(\Sigma) \equiv k(\Sigma) \frac{\tau^*_{cr}}{\vert \tau^*(\Sigma) \vert}
\end{equation}
where $k(\Sigma)$ is a dimensionless scalar. In this general formulation, $r_{SE}$ is a function of the macroscopic stress level $\Sigma$. By choosing an appropriate $k(\Sigma)$ this functional dependence can be eliminated. Furthermore, the appropriate choice of $k(\Sigma)$ will allow $r_{SE}$ to be cast as a ratio of strength to effective stiffness. Substituting Equation~\ref{eqn:y2e_StressRatio} into Equation~\ref{eqn:y2e_Y2E} yields
\begin{equation}\label{eqn:y2e_kDeriv1}
r_{SE}(\Sigma) = k(\Sigma) \frac{\Vert \tensSym{\sigma}^\prime_o \Vert}{\Vert \tensSym{\sigma}^\prime(\Sigma)\Vert}
\end{equation}
Anisotropic Hooke's law relates the local stress and local elastic strain $\tensSym{\epsilon}$
\begin{equation}
\label{eqn:y2e_HookeLocal}
\tensSym{\sigma} = \mathbf{C} \tensSym{\epsilon}
\end{equation}
where $\mathbf{C}$ is the fourth-order stiffness tensor\footnote{This form of Hooke's Law differs from that presented in 
Section~\ref{sec:model_methods} (Equations~\ref{eq:cauchy_kirchhoff} -- \ref{eq:hooke_kirchhoff_form_dev}) 
in that here the Cauchy stress appears and there the Kirchhoff stress is used.  The difference is small between the two, generally a fraction of a percent.  Using the Cauchy stress here simplifies the development without altering the results significantly.}.
Substituting Equation~\ref{eqn:y2e_HookeLocal} into Equation~\ref{eqn:y2e_kDeriv1} yields
\begin{equation}
\label{eqn:y2e_kDeriv2}
r_{SE}(\Sigma) = k(\Sigma) \frac{\Vert \tensSym{\sigma}^\prime_o \Vert}{\Vert \mathbf{C} \tensSym{\epsilon}^\prime (\Sigma)\Vert}
\end{equation}
Choosing $k(\Sigma) = E_\mathit{eff}(\Sigma)$ where $E_\mathit{eff}$ is the effective macroscopic strain yields
\begin{equation}\label{eqn:y2e_kDeriv3}
r_{SE}(\Sigma) = \frac{\Vert \tensSym{\sigma}^\prime_o \Vert}{\Vert \mathbf{C} \tensSym{\epsilon}^\prime (\Sigma) / E_\mathit{eff}(\Sigma)\Vert}
\end{equation}
The ratio $\tensSym{\epsilon}^\prime (\Sigma) / E_\mathit{eff}(\Sigma)$ is independent of $\Sigma$. The strength-to-stiffness ratio in this formulation is likewise independent of the stress coefficient $\Sigma$. It still depends on the principal directions of the macroscopic stress and the ratios of principal macroscopic stress components, contained in the stress basis tensor $\hat{\tensSym{\Sigma}}$. The strength-to-stiffness ratio can therefore be written as
\begin{equation}\label{eqn:y2e_kDeriv4}
r_{SE} = \frac{\Vert \tensSym{\sigma}^\prime_o \Vert}{\Vert \mathbf{C} \tensSym{\epsilon}^\prime / E_\mathit{eff} \Vert}
\end{equation}
In this form, it is clear that $r_{SE}$ represents the ratio of an effective strength $\Vert \tensSym{\sigma}^\prime_o \Vert$ to an effective stiffness (stress per unit strain) $\Vert \mathbf{C} \tensSym{\epsilon}^\prime / E_\mathit{eff} \Vert$. The strength-to-stiffness ratio $r_{SE}$ is a dimensionless quantity. 

In practice, the form of Equation~\ref{eqn:y2e_Y2E} is used to calculate strength-to-stiffness, with $k = E_\mathit{eff}$
\begin{equation}\label{eqn:y2e_Y2EEval}
r_{SE} = E_\mathit{eff} \frac{\tau^*_{cr}}{\vert \tau^* \vert}
\end{equation}
The variables appearing in Equation~\ref{eqn:y2e_Y2EEval}, $E_\mathit{eff}$ and $\tau^*$, are evaluated at any macroscopic stress $\tensSym{\Sigma}$ in the elastic regime that satisfies Equation~\ref{eqn:y2e_MonotonicLoad} for the load path of interest. 
The local deviatoric elastic strain $\tensSym{\epsilon}^\prime$ can either be evaluated from one load increment of a purely elastic finite element simulation or approximated with an isostrain assumption. 

%% file: Model_Methods.tex
The strength-to-stiffness parmeter is developed in the context of a
mathematical model of the elastoplastic behavior of polycrystalline solids.   This model is set at the physical length scale of crystals such that a entire body is an aggregate of grains, numbering from just a few to possible millions.  A material point within the model is a domain within a crystal in which the properties are those of subdomain of a single crystal.  A brief explanation of the constitutive equations of the model is presented in this section.  These equations are part of a complete mechanical model for the motion of a deforming polycrystal.  
The complete model is outlined in the appendix.  For more detailed information the reader is referred to~\cite{Dawson14a, Marin98a, Marin98b}.

The constitutive model for a volume of material within one grain of deforming polycrystal employs a kinematic decomposition of the deformation into a sequence of deformations due to crystallographic slip, rotation, and elastic stretch. Using this decomposition, the deformation gradient, $\tensSym{F}$, can be represented as
\begin{equation}\label{eqn:kinematic_decomp}
\tensSym{F} = \tensSym{V}^e  \tensSym{R}^*  \tensSym{F}^p 
\end{equation}
where $\tensSym{F}^p$, $\tensSym{R}^*$, and  $\tensSym{V}^e$ correspond to crystallographic slip, rotation, and elastic stretch, respectively. A schematic of this decomposition is provided in~\cite{Dawson14a}. This decomposition defines a reference configuration $\mathbf{B}_0$, a deformed configuration $\mathbf{B}$, and two intermediate configurations $\bar{\mathbf{B}}$ and $\hat{\mathbf{B}}$. The state equations are written in the intermediate $\hat{\mathbf{B}}$ configuration defined by the relaxation of the elastic deformation from the current $\mathbf{B}$ configuration. Elastic strains, $\tensSym{\epsilon}^e$, are required to be small, which allows the elastic stretch tensor to be written as
\begin{equation}
\tensSym{V}^e = \tensSym{I} + \tensSym{\epsilon}^e
\end{equation}
The velocity gradient, $\tensSym{L}$, is calculated from the deformation gradient using the relationship
\begin{equation}
\tensSym{L} = \dot{\tensSym{F}} \tensSym{F}^{-1}
\end{equation}
The velocity gradient can be decomposed into a symmetric deformation rate, $\tensSym{D}$, and skew-symmetric spin rate, $\tensSym{W}$
\begin{equation}
\tensSym{L} = \tensSym{D} + \tensSym{W}
\end{equation}
A generic symmetric tensor, $\tensSym{A}$, can be additively decomposed into mean and deviatoric components
\begin{equation}\label{eqn:vol_dev_decomp}
\tensSym{A} = \sfrac{1}{3} \mathrm{tr} (\tensSym{A}) \tensSym{I} + \tensSym{A}^\prime
\end{equation}
where $\sfrac{1}{3} \mathrm{tr} (\tensSym{A})$ is the scalar mean component and $\tensSym{A}^\prime$ is the tensorial deviatoric component.
Utilizing Equations~\ref{eqn:kinematic_decomp}-\ref{eqn:vol_dev_decomp} the volumetric deformation rate, deviatoric deformation rate, and spin rate can be expressed as
\begin{equation}
\mathrm{tr} \left( \tensSym{D} \right) = \mathrm{tr} \left( \dot{\tensSym{\epsilon}}^e \right)
\label{eq:kinematic_decomp_meandefrate}
\end{equation}
\begin{equation}
\tensSym{D}^\prime = \dot{\tensSym{\epsilon}}^{e\prime} + \hat{\tensSym{D}}^p + \tensSym{\epsilon}^{e\prime} \cdot \hat{\tensSym{W}}^p - \hat{\tensSym{W}}^p \cdot \tensSym{\epsilon}^{e\prime}
\label{eq:kinematic_decomp_devdefrate}
\end{equation}
\begin{equation}
\tensSym{W} = \hat{\tensSym{W}}^p + \tensSym{\epsilon}^{e\prime} \cdot \hat{\tensSym{D}}^p - \hat{\tensSym{D}}^p \cdot \tensSym{\epsilon}^{e\prime}
\label{eq:kinematic_decomp_spin}
\end{equation}
where $\hat{\tensSym{D}}^p$ and $\hat{\tensSym{W}}^p$ are the plastic deformation and spin rates.

Constitutive equations relate the stress to the deformation. The Kirchhoff stress, $\tensSym{\tau}$, in the $\hat{\mathbf{B}}$ configuration is related to the Cauchy stress, $\tensSym{\sigma}$, in the current configuration $\mathbf{B}$ by the determinant of the elastic stretch tensor
\begin{equation}
\tensSym{\tau} = \mathrm{det} (\tensSym{V^e}) \tensSym{\sigma}
\label{eq:cauchy_kirchhoff}
\end{equation}
The Kirchhoff stress is related to the elastic strain through anisotropic Hooke's law
\begin{equation}
\mathrm{tr} \left( \tensSym{\tau} \right) = 3K \mathrm{tr} \left( \tensSym{\epsilon}^e \right)
\label{eq:hooke_kirchhoff_form_mean}
\end{equation}
\begin{equation}
\tensSym{\tau}^\prime = \mathbf{C} \tensSym{\epsilon}^{e\prime}
\label{eq:hooke_kirchhoff_form_dev}
\end{equation}
where $K$ is the bulk modulus and $\mathbf{C}$ is the fourth-order stiffness tensor.

Plastic deformation due to crystallographic slip occurs on a restricted set of slip systems. For FCC crystals, slip occurs on the \{111\} planes in the [110] directions. For BCC crystals, slip on the \{110\} planes in the [111] directions is considered. The plastic deformation rate and plastic spin rate are given by 
\begin{equation}
\hat{\tensSym{D}}^p = \sum_\alpha \dot{\gamma}^\alpha \hat{\tensSym{P}}^\alpha
\end{equation}
\begin{equation}
\hat{\tensSym{W}}^p = \dot{\tensSym{R}^*} \tensSym{R}^{*T} + \sum_\alpha \dot{\gamma}^\alpha \hat{\tensSym{Q}}^\alpha
\end{equation}
where $\hat{\tensSym{P}}^\alpha$ and $\hat{\tensSym{Q}}^\alpha$ are the symmetric and skew symmetric components of the Schmid tensor $\hat{\tensSym{T}}^\alpha$, and $\dot{\gamma}^\alpha$ is the shear rate on the $\alpha$-slip system. The Schmid tensor is defined as the dyad of the slip direction, $\hat{\mathbf{s}}^\alpha$, and slip plane normal, $\hat{\mathbf{m}}^\alpha$
\begin{equation}
\hat{\tensSym{T}}^\alpha = \hat{\mathbf{s}}^\alpha \otimes \hat{\mathbf{m}}^\alpha
\end{equation}
The slip system shear rate for a given slip system is related to the critical resolved shear stress on that slip system, $\tau^\alpha$, by a power law relationship
\begin{equation}
\dot{\gamma}^\alpha = \dot{\gamma}_0 \left( \frac{\vert \tau^\alpha \vert}{g^\alpha} \right)^\frac{1}{m} \mathrm{sgn} \left( \tau^\alpha \right)
\end{equation}
where $\dot{\gamma}_0$ is a reference slip system shear rate and $g^\alpha$ is the slip system strength. The resolved shear stress is the projection of the deviatoric stress onto the slip system
\begin{equation}
\tau^\alpha = \tensSym{\tau}^\prime : \hat{\tensSym{P}}^\alpha
\label{eq:resolved_shear_stress}
\end{equation}

A modified Voce hardening law is used to describe slip system strength evolution 
\begin{equation} \label{eqn:Voce}
\dot{g}^\alpha = h_0 \left( \frac{g_s - g^\alpha}{g_s - g_0} \right)^{n^\prime} \sum_\alpha \dot{\gamma}^\alpha
\end{equation}
where $h_0$ is the reference hardening rate, $g_s$ is the saturation strength, and $n^\prime$ is the hardening exponent. The hardening law is isotropic; at a given material point, all slip systems harden at the same rate. 

 The rate of lattice re-orientation follows directly from
the equation for the plastic spin, assuming that the slip system shearing rates are known.  
Written in terms of the Rodrigues vector:
\begin{equation}
\dot\rodvec = \frac{1}{2} (\spinvec + (\spinvec \cdot \rodvec) \rodvec + \spinvec \times \rodvec)
\label{eq:rod_rate}
\end{equation}
where
\begin{equation}
\spinvec = \vectop\left( \pxspinhat - \sumss \gammadot \skwschmid \right)
\end{equation}

%% file: material_and_virtualpolycrystal.tex
\begin{figure}[htp]
\centering
{\includegraphics[width=8.5cm]{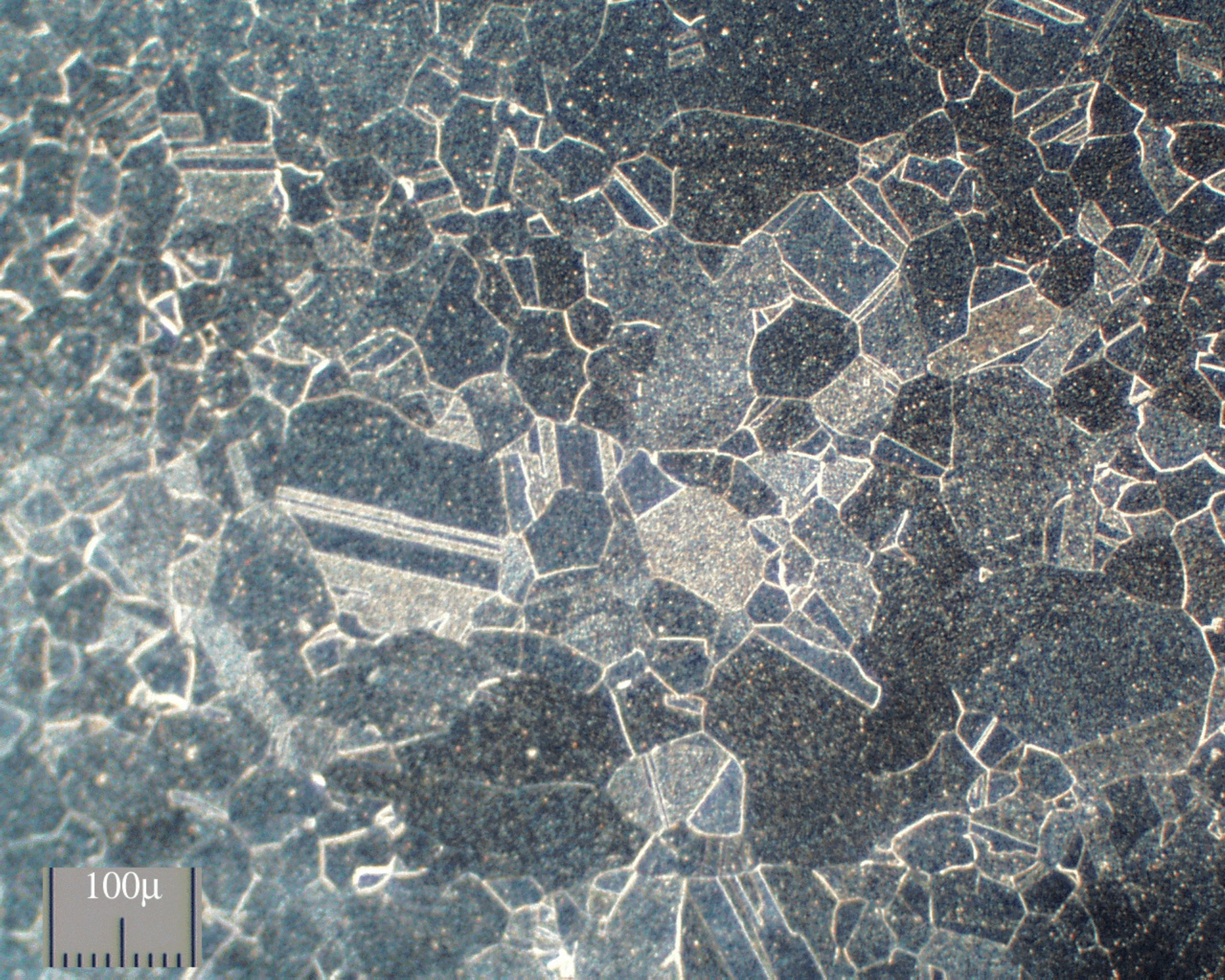} }
\caption{Optical micrograph of AL6XN microstructure.}
\label{fig:al6xn-micros}
\end{figure}

The multiaxial strength-to-stiffness framework is used to examine the initiation and propagation of yielding in AL6XN. 
Al6XN is a super-austenitic stainless
steel with  a nominal composition 
consisting of 49 Wt\% Fe, 20 Wt\% Cr, 24 Wt\% Ni, 6 Wt\% Mo, and
$<1$ Wt\% each of N, Mn, Si, and Cu (as per fact sheet by \cite{SS1}). 
It has a face-centered cubic (FCC) crystal structure.
A micrograph, given in Figure~\ref{fig:al6xn-micros}
for a specimen etched to highlight the grain boundaries
is typical of this alloy.
The structure is essentially single-phase with a relatively equi-axed grains.  
The average grain size is on the order of 50 $\mu$m.
There is evidence of mechanical twins from prior deformation and of second phase particles (referred to as the sigma phase) that are much smaller
than a typical grain of the primary phase. 

\begin{figure}[p]
    \centering
    \includegraphics[width = 0.6\textwidth]{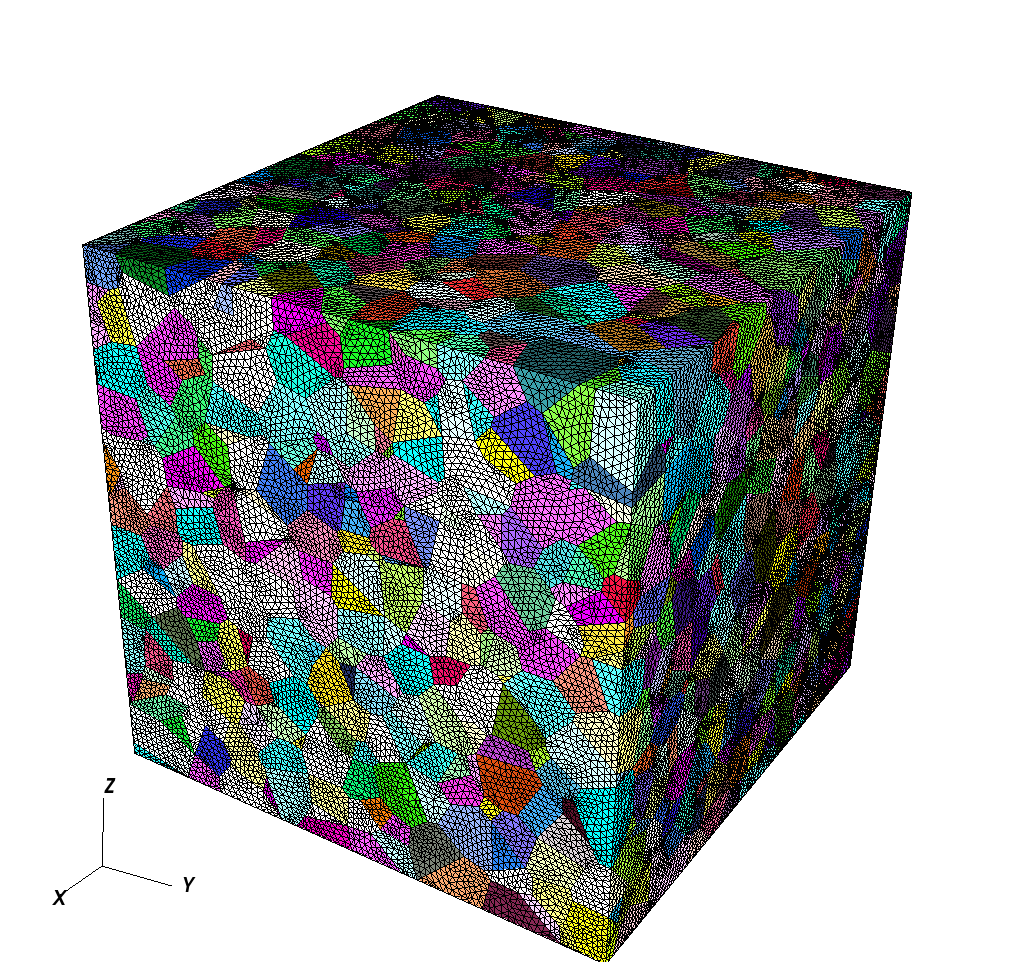}
    \caption{Virtual microstructure for AL6XN. The virtual polycrystal is comprised of 5,000 grains and discritized by over 550,000 finite elements.}
    \label{fig:microstructure}
\end{figure}

To perform the finite element simulations of the mechanical response, a virtual polycrystal, consisting of 5,000 crystals was instantiated, as depicted in Figure~\ref{fig:microstructure}. The virtual polycrystal was discretized using finite elements. The finite element mesh 
was created using the Neper code~\cite{quey_large-scale_2011} and  contains 550,680 elements, or an average of about 110 elements per crystal.

The measured crystallographic texture of this sample is characteristic of a weak rolling texture.  For the purpose of demonstrating  the multiaxial strength-to-stiffness parameter, however, the texture was taken to be uniform (equal probability of all orientations, often referred to as 'random') as we did not want to bias the results toward one particular texture. 
Thus, to initialize the texture of the virtual polycrystal, orientations were chosen randomly from a uniform distribution and assigned to grains.  All of the finite elements within a grain were given the same initial orientation.  Evolving the orientations with plastic deformation was carried out element-by-element according to Equation~\ref{eq:rod_rate}.

\begin{figure}[p]
    \centering
    \includegraphics[scale = 0.4]{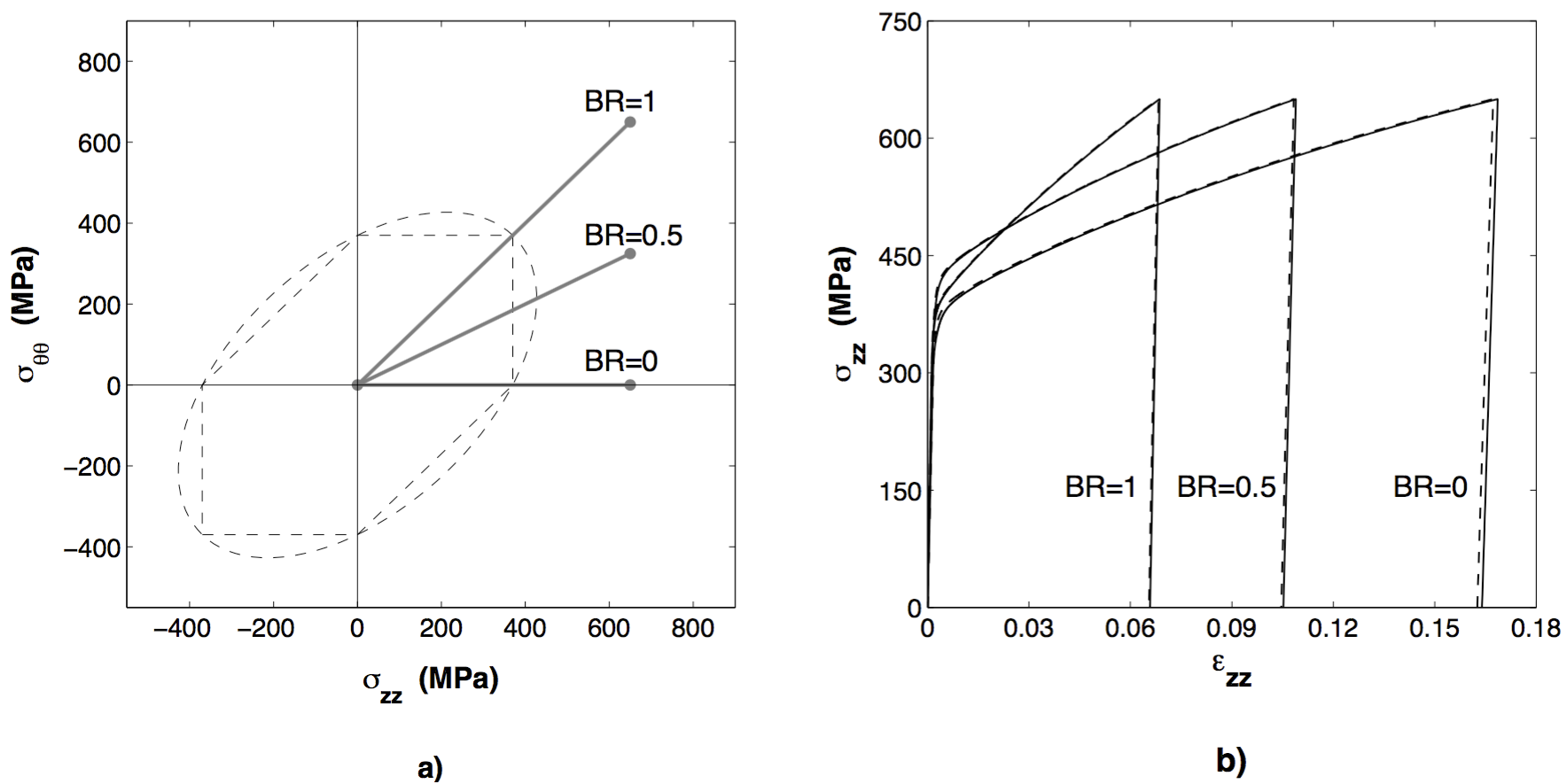}
    \caption{Reproduced Figure 3 of \cite{Marin12a} showing the simulated stress-strain responses for three biaxial stress ratios at both low (dashed) and high (solid) elastic anisotropy.}
    \label{fig:stress-strain}
\end{figure}

The same elastic and plastic crystal properties as have been used in prior modeling efforts related to this alloy~\cite{Marin08a,Marin12a}. The elastic moduli are take from \cite{Ledbetter_1984}, and are listed in Table~\ref{tab:elasticmoduli}.  The plasticity parameters, listed in Table~\ref{tab:plasticityparameters},  were determined by fitting experimental records of tensile tests on this alloy.
  Simulated responses are shown in Figure~\ref{fig:stress-strain}.

\medskip
\begin{table}[ht]
\centering
  \begin{tabular}{|| c | c | c |  c  | c ||}
  \hline
  \hline
   $C_{11}$ & $C_{12}$  & $C_{44}$ &$r_E$ & $A$\\ [0.5ex]
  \hline
  \hline
204.6 & 137.7 & 126.2 & 3.2  &  3.77  \\
  \hline
  \end{tabular}
\caption{Elastic moduli (GPa) and anisotropy metrics for \protect{AXL6N}~\cite{Ledbetter_1984}.}
\label{tab:elasticmoduli}
\end{table}
\medskip
 \begin{table}[h]
\centering
\begin{tabular}{||c|c|c|c|c|c||}
\hline
\hline
 $\gdotz$ (${{\rm s}^{-1}}$) & $m$ & $h_{0}$ (MPa) & $g_0$ (MPa)& $n^\prime$  & $g_s$ (MPa)  \\
\hline
 1.0 & 0.02 & 375 & 160 & 1 & 1000  \\
\hline
\hline
\end{tabular}
\caption{Plastic parameters for \protect{AXL6N}~\cite{Marin12a}.  See Section~\ref{sec:model_methods} for parameter definitions.}
\label{tab:plasticityparameters}
\end{table}
\medskip

Symmetry boundary conditions were applied to the three orthogonal surfaces described by $x=0$, $y=0$, and $z=0$. Symmetry boundary conditions consist of zero velocity component in the the direction normal to the surface and zero traction component tangent to the surface. 

Mixed velocity/traction boundary conditions were applied to the other three orthogonal surfaces. On each surface, the normal component of velocity was prescribed uniformly over the surface, and the tangenetal component of traction was set to zero. The normal velocity on the $z$-surface was constant throughout the simulation to produce a constant engineering strain rate of $10^{-4} \,{\rm s}^{-1}$ in the axial $z$-direction. Iterations were performed on the normal components of velocity prescribed on the other two surfaces at each step of the simulation, so as to maintain the prescribed ratios of macroscopic stress components. 
The prescribed ratios were chosen to give particular biaxial stress states ranging from simple uniaxial stress to balanced biaxial stress.   The stress tensor for these stress states can be written in the principal bases as:
\begin{equation}
\left[ \cauchy \right]
= \left[  \begin{array}{ccc}  \sigma_{xx} & 0 & 0 \\ 0 &  \sigma_{yy}  (=0)  & 0 \\ 0 & 0 & \sigma_{zz} \end{array} \right]
 \label{eqn:biaxial_stress_state}
\end{equation}
By adjusting the boundary velocities as described above, two components of the stress, $\sigma_{xx}$ and $\sigma_{zz}$, are controlled to give prescribed levels of stress biaxiality. 
The remaining component of the stress, $\sigma_{yy}$, 
is held fixed at zero. 
The level of stress biaxiality is quantified by the biaxial ratio (BR): $\frac{\sigma_{xx}}{\sigma_{zz}}$.
For uniaxial tension,  the BR equals zero, while for balance biaxial tension, the BR is unity.  
Simulations were conducted for five biaxial ratios: 0.0, 0.25, 0.5, 0.75 and 1.0.

%% file: Y2E_Application.tex
The strength-to-stiffness ratio is a useful metric for predicting the progression of yielding within a polycrystalline aggregate because is incorporates the combined effects of the stiffness, which regulates the relative stress level among crystals sharing a common load, and the  directional strength, which limits the stress that may be carried by a given strength.  To illustrate this point, the mechanical responses are examined for two virtual samples with low and high elastic anisotropy. The case of low elastic anisotropy is actually the isotropic limit.  The single crystal properties for the polycrystal with high elastic anisotropy are those of AL6XN for both elastic and plastic moduli.  For the elastically isotropic case, the same plastic moduli are used but the elastic moduli are re-defined to correspond to isotropic behavior.    
For each of the polycrystals, the five levels of biaxial ratio are examined with through correlatons between the volume fraction of elements that have reached yielding and three metrics related to the mechanical behavior: the Schmid factor, the Taylor factor and the strength-to-stiffness ratio.  

\begin{figure}[ht!]
    \centering
    \subfigure[Reciprocal Schmid factor]
    {\includegraphics[trim = 0in 0in 0.0in 0in, clip, scale = 0.7]{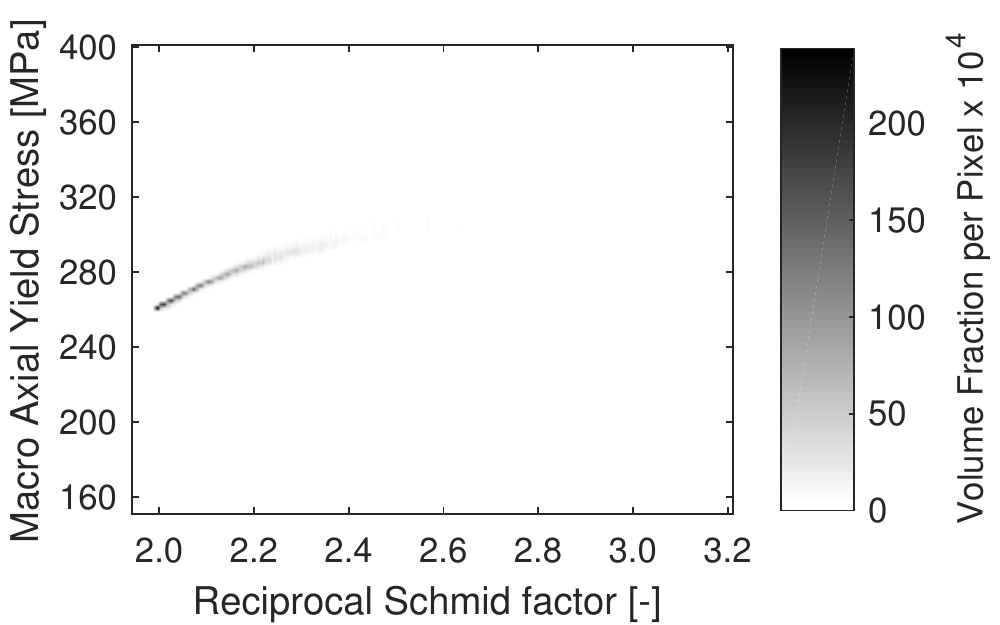}
    \label{fig:YieldStressSchmidIsotropic}}
    \subfigure[Taylor factor]
    {\includegraphics[trim = 0in 0in 0.0in 0in, clip, scale = 0.7]{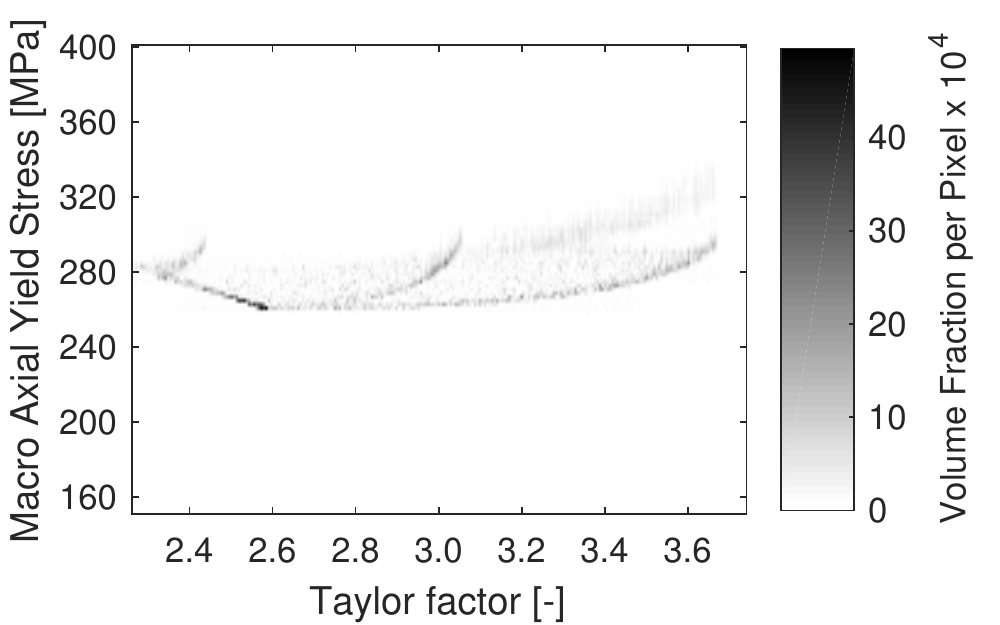}
    \label{fig:YieldStressTaylorIsotropic}}
    \subfigure[Strength-to-stiffness (isostrain)]
    {\includegraphics[trim = 0in 0in 0.0in 0in, clip, scale = 0.7]{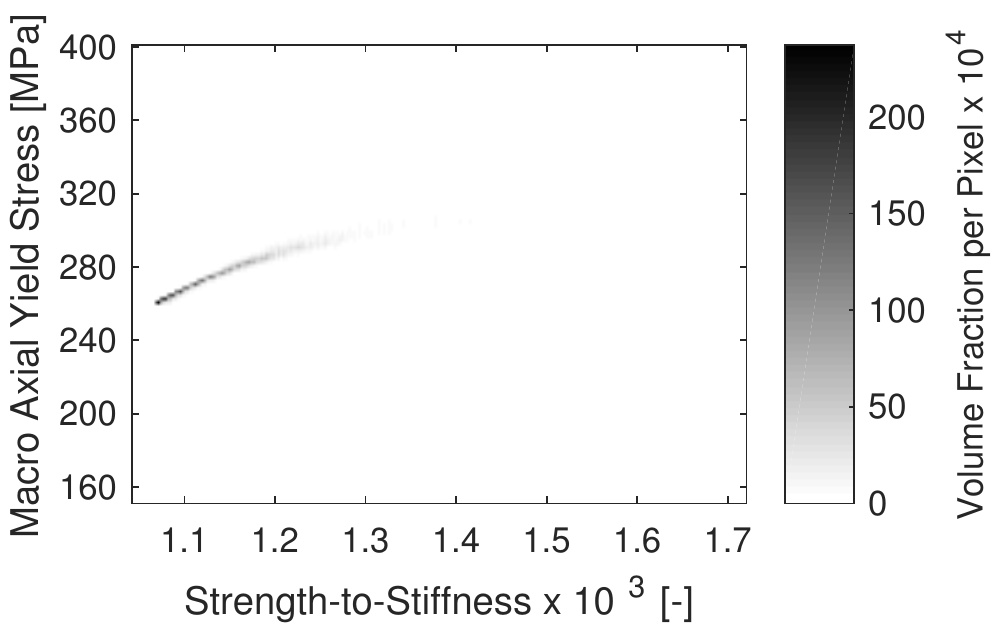}
    \label{fig:YieldStressY2EIsoIsotropic}}
    \subfigure[Strength-to-stiffness]
    {\includegraphics[trim = 0in 0in 0.0in 0in, clip, scale = 0.7]{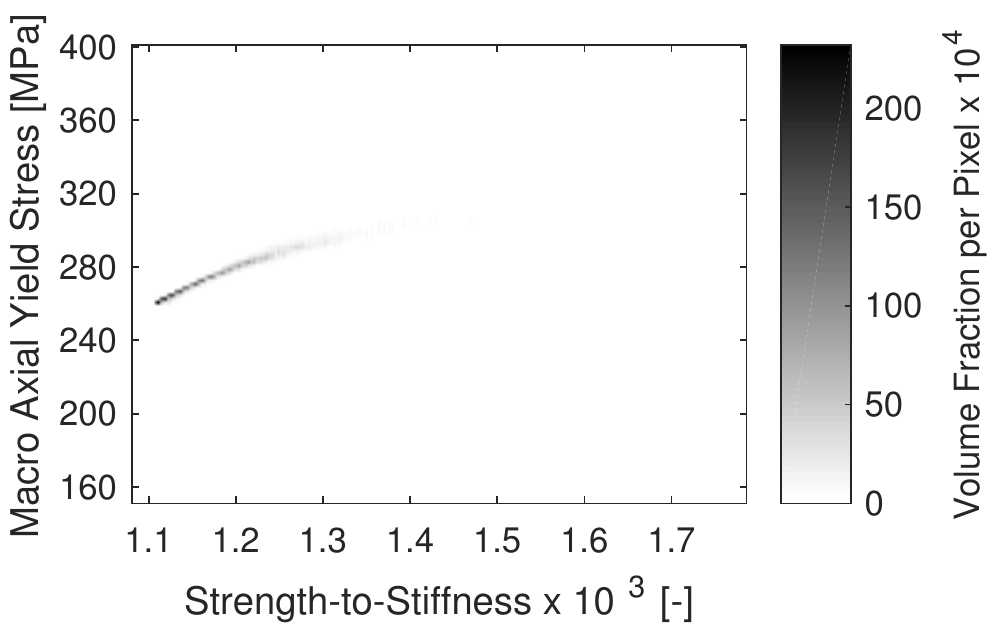}
    \label{fig:YieldStressY2EIsotropic}}
    \caption{Correlation between elemental yield stress and strength and strength-to-stiffness measures for an elastically isotropic material under uniaxial loading. In the special case of elastic isotropy at the microscale, the strength-to-stiffness parameter is dependent only on anisotropic strength. For this special case, the Schmid and strength-to-stiffness parameters both corralate to the macroscopic strength at which localized yielding occurs.}  
    \label{fig:ElemYieldStressIsotropic}   
\end{figure}

To illustrate that the anisotropies of both stiffness and strength are important in quantifying yielding in polycrystals, we compare the correlations for the low and high elastic anisotropy behaviors under uniaxial loading (BR=0). 
Beginning with the low elastic anisotropy case,
binned scatter plots are presented in  Figure~\ref{fig:ElemYieldStressIsotropic} for the correlations between the macroscopic stress at which elements yield and the three metrics. 
In these plots, data are binned according to each of the three metrics  and macroscopic elemental yield stress.  Intensity corresponds to the volume fraction of the aggregate contained in each bin.
Several trends are present in the correlations.
 First, note that if the elastic response is isotropic, then there is no difference between the correlation made with the Schmid factor and the correlation made with the strength-to-stiffness parameter.  This is expected since the directional stiffness is a constant for isotropic behavior.  Next, note that the correlation is the same for the results generated with an isostrain assumption and with the finite element model.  This implies that the stress is uniform over the polycrystal in both cases, which is true since the crystal elastic properties are uniform when the elasticity is isotropic.  For crystalline solids, isotropy is the exception, not the rule.  Finally, note the structure of the correlation with the Taylor factor.  Several arcs are evident among the scatter of points.  These are associated with the topology of the single crystal yield surface which has five distinct families of vertices.

\begin{figure}[ht!]
    \centering
    \subfigure[Reciprocal Schmid factor]
    {\includegraphics[trim = 0in 0in 0.0in 0in, clip, scale = 0.7]{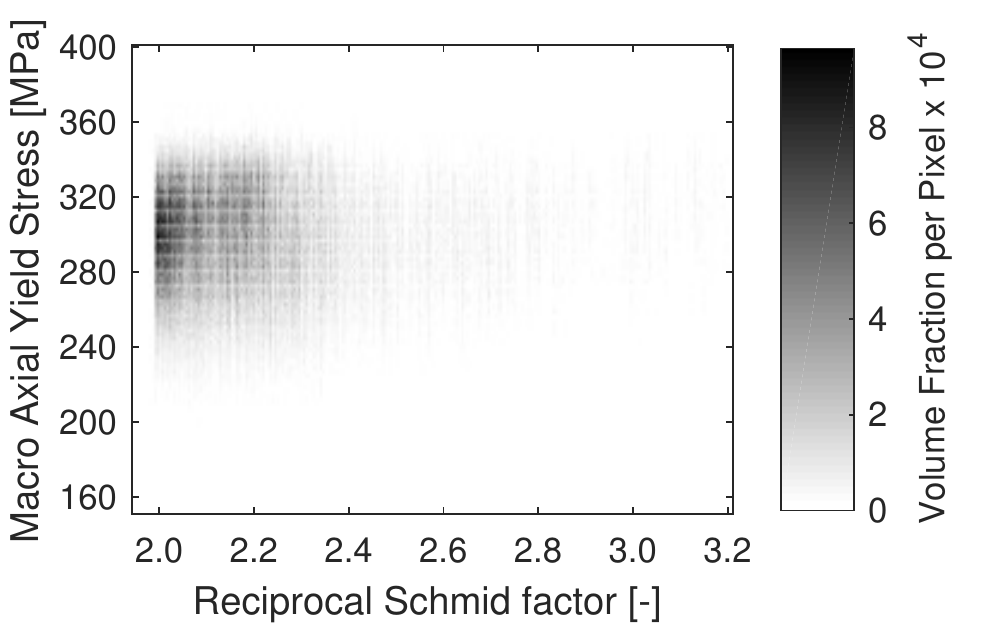}
    \label{fig:YieldStressSchmid}}
    \subfigure[Taylor factor]
    {\includegraphics[trim = 0in 0in 0.0in 0in, clip, scale = 0.7]{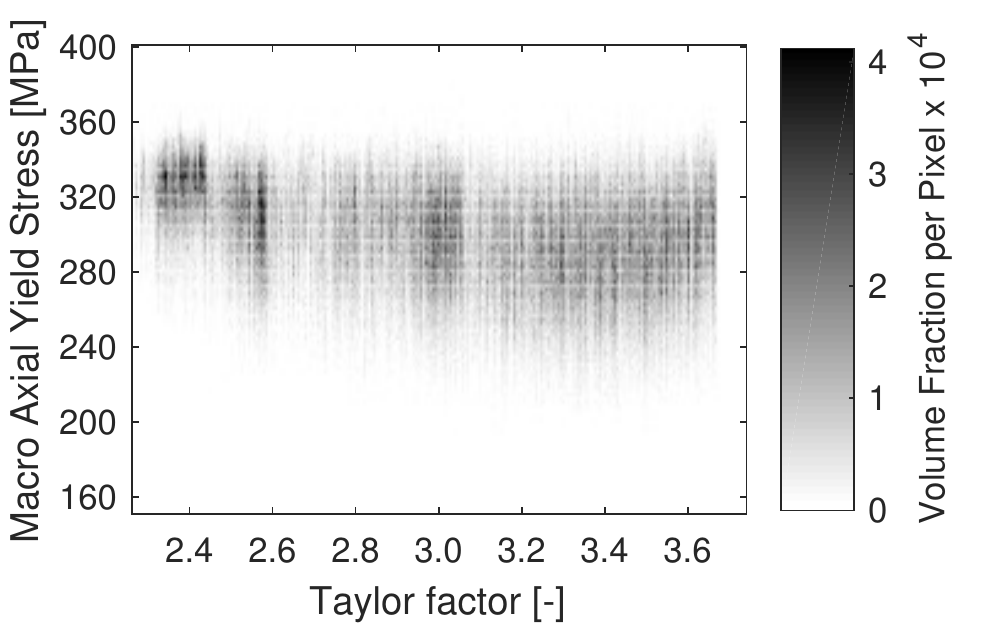}
    \label{fig:YieldStressTaylor}}
    \subfigure[Strength-to-stiffness (isostrain)]
    {\includegraphics[trim = 0in 0in 0.0in 0in, clip, scale = 0.7]{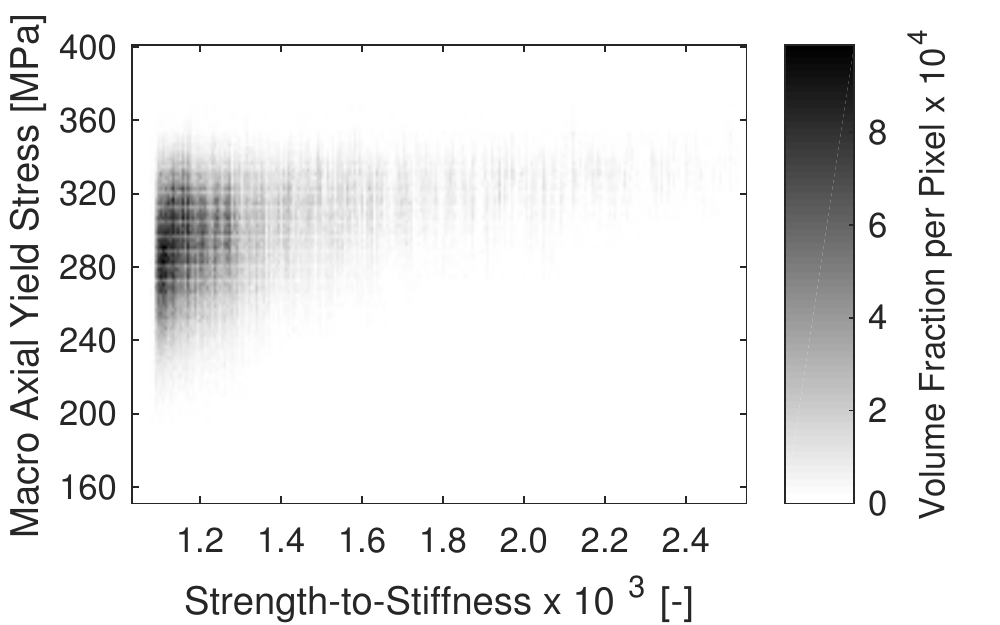}
    \label{fig:YieldStressY2EIso}}
    \subfigure[Strength-to-stiffness]
    {\includegraphics[trim = 0in 0in 0.0in 0in, clip, scale = 0.7]{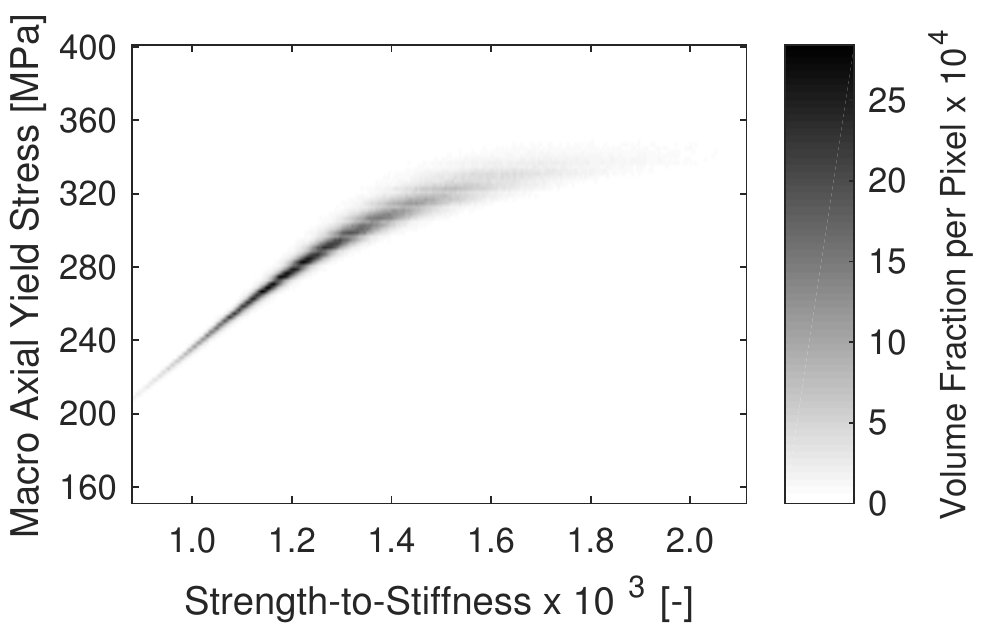}
    \label{fig:YieldStressY2E}}
    \caption{Correlation between elemental yield stress and strength and strength-to-stiffness measures for AL6XN under uniaxial loading. Both strength and stiffness are important in governing yield behavior. The strength-to-stiffness parameter, which incorporates neighborhood effects, captures the relative order in which elements yield.}
    \label{fig:ElemYieldStress}
\end{figure}

The same plots shown the low elastic anisotropy case are
now presented for the high elastic anisotropy case in Figure~\ref{fig:ElemYieldStress}.  Here, both the elastic and plastic crystal properties are for AL6XN. 
The contrasts between the two sets of figures are striking.
There is no correlation between the Schmid and Taylor factors and macroscopic elemental yield stress (Figures~\ref{fig:YieldStressSchmid} and~\subref{fig:YieldStressTaylor}). The vertical streaking is due to the fact that, although the strength parameter is constant over the grain, the entirety of a grain does not yield at the same macroscopic stress. The heavy streaks are likely produced by large BCC grains. It is also interesting to note that there is a pocket of elements with low Taylor factors that yield at a higher macroscopic stress than the other elements. This pocket corresponds to the $\{100\}||[001]$ crystallographic fiber, which has low strength, but also low directional stiffness. These crystals have relatively high strength-to-stiffness and therefore yield later than other crystals. Strength-to-stiffness formulated with an isostrain assumption exhibits some correlation with macroscopic elemental yield stress (Figure~\ref{fig:YieldStressY2EIso}). The elements that are first to yield all have low strength-to-stiffness, and elements with high strength-to-stiffness are among the last to yield. However, there is a large volume fraction of material with low strength-to-stiffness that yields over a range of 100~MPa. The isostrain approximation of strength-to-stiffness is not very predictive. It does not take into account that the local stress is different from the macroscopic stress due to intergranular interactions produced by compatibility constraints. Only when the local stress, or equivalently the local elastic strain, from the finite element simulation is incorporated into the strength-to-stiffness formulation is there a strong correlation between strength-to-stiffness and macroscopic elemental yield stress (Figure~\ref{fig:YieldStressY2E}). This comparison highlights the importance of neighborhood effects and illustrates that both strength and stiffness are important in governing the initiation and propagation of yielding.

The comparisons of the two sets of correlations overall show that the presence of elastic anisotropy leads to spatial heterogeneity in the stress distribution that must be considered when determining if the stress is sufficient to initiate yielding.  The strength alone is insufficient as it provides no information on the heterogenity of the stress within the elastic domain and thus is insufficent to determine how close the stress lies to the yield surface.

\begin{figure}[p]
    \centering
    \subfigure[$BR = 0.00$]
    {\includegraphics[trim = 0in 0in 0.9in 0in, clip, scale = 0.6]{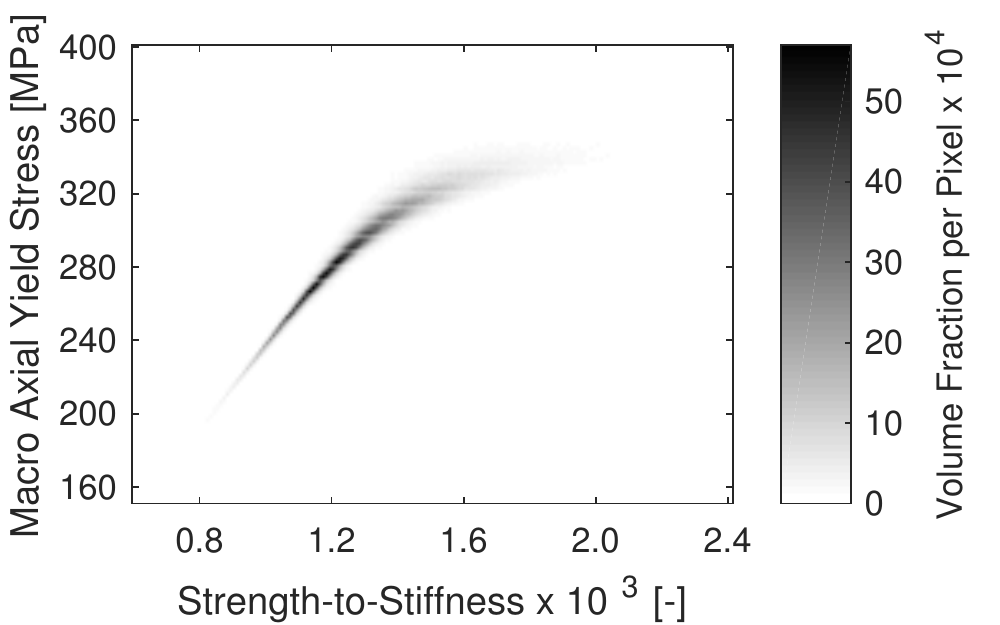}
    \label{fig:Y2E-BR000}} 
    \subfigure[$BR = 0.25$]
    {\includegraphics[trim = 0in 0in 0.9in 0in, clip, scale = 0.6]{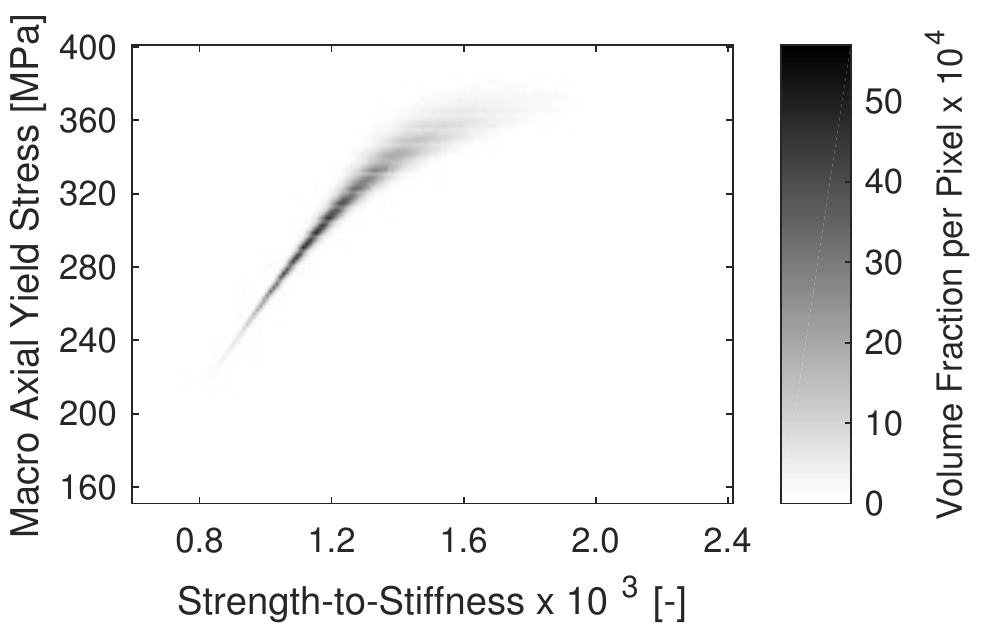}
    \label{fig:Y2E-BR025}} 
    \subfigure[$BR = 0.50$]
    {\includegraphics[trim = 0in 0in 0.9in 0in, clip, scale = 0.6]{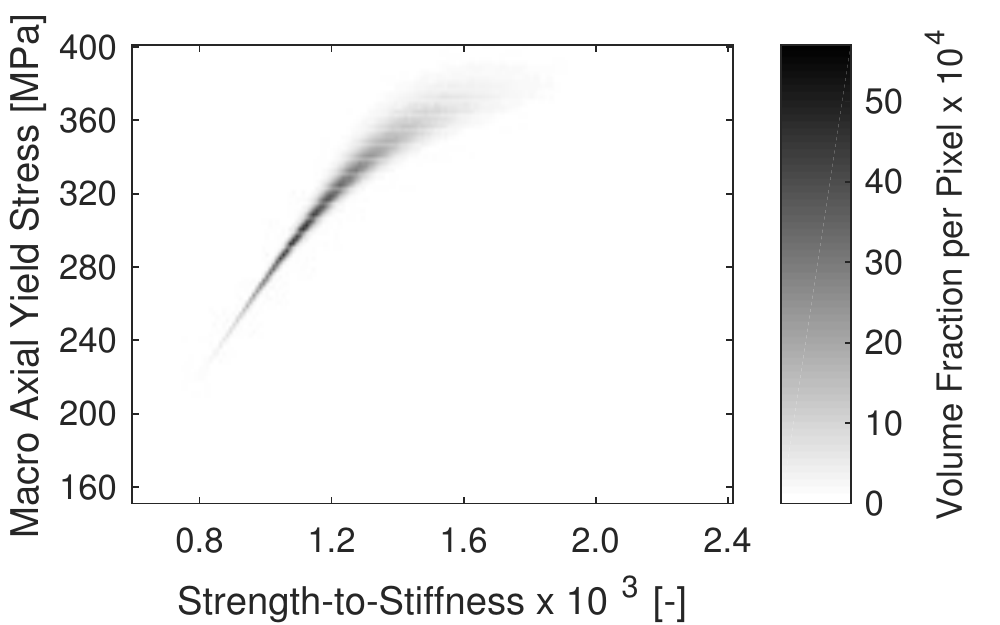}
    \label{fig:Y2E-BR050}} 
    \subfigure[$BR = 0.75$]
    {\includegraphics[trim = 0in 0in 0.9in 0in, clip, scale = 0.6]{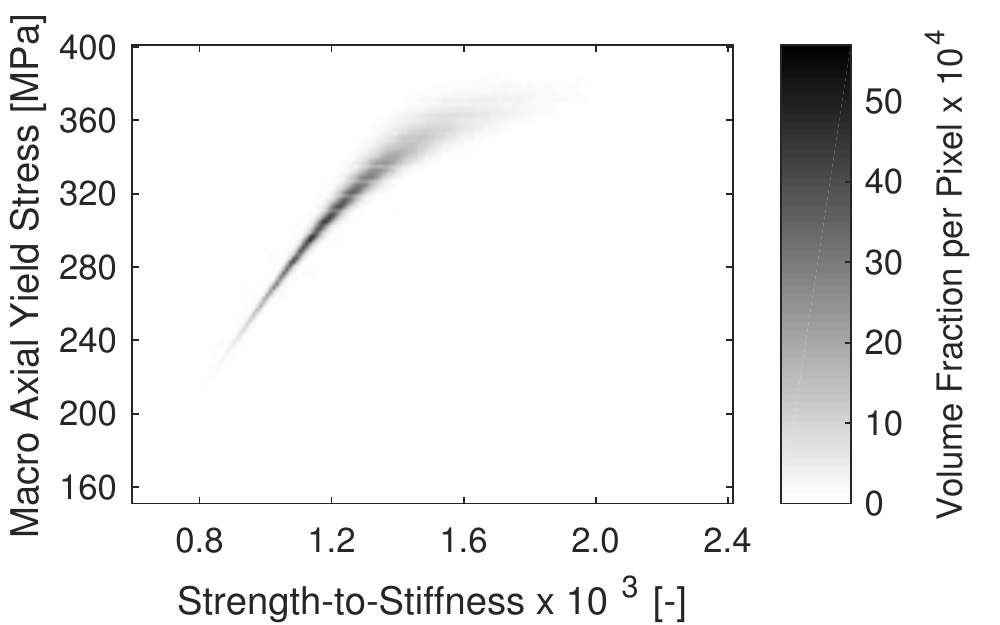}
    \label{fig:Y2E-BR075}} 
    \subfigure[$BR = 1.00$]
    {\includegraphics[trim = 0in 0in 0.9in 0in, clip, scale = 0.6]{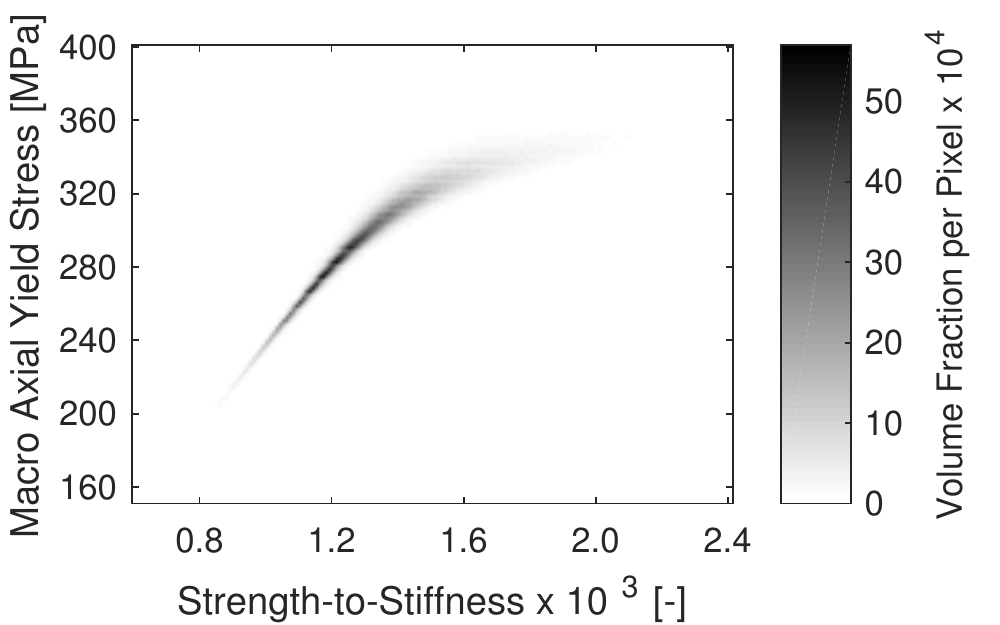}
    \label{fig:Y2E-BR100}} 
    \caption{The multiaxial strength-to-stiffness ratio governs the order in which elements yield. Elements with low strength-to-stiffness yield before elements with high strength-to-stiffness.}\label{fig:Y2E}
\end{figure}

We now examine how the correlations between the macroscopic stress at which elements yield and strength-to-stiffness for the full range of stress biaxiality.
Binned scatter plots between strength-to-stiffness and the macroscopic stress at which elements yield are presented in Figure~\ref{fig:Y2E} for five levels of stress biaxiality. Again, data are binned according to both strength-to-stiffness and macroscopic elemental yield stress and the plotted intensity corresponds to the volume fraction of the aggregate contained in each bin. There is a strong, nonlinear correlation between strength-to-stiffness and macroscopic elemental yield stress for all biaxial stress states. The downward concavity of the curves is due to the increase in the local load increment, relative to the macroscopic load increment, that occurs to elastic elements when other elements yield. When an element yields, its ability to carry additional load is significantly reduced. Additional incremental load must be carry by the remaining elastic elements, and so the effective load increment for the elastic elements increases. As the local load increment increases relative to the macroscopic load increment, elements yield at lower macroscopic stresses than if the local load increment were constant, producing the downward curvature observed in the figures. The correlation between strength-to-stiffness and macroscopic elemental yield stress is stronger at low strength-to-stiffness than at high strength-to-stiffness. The correlation decreases over the course of the elasto-plastic transition because the analysis is based on a linearization of behavior in the elastic regime. As yielding progresses, local stresses evolve, deviating from the linearized values. The correlation is therefore stronger for elements that yield earlier in the elasto-plastic transition.

%% file: Yield_Predict_Formulation.tex
Analysis similar to strength-to-stiffness can be used to predict the macroscopic stress at which a region will yield. The same assumptions about the macroscopic and local load histories that were used for the strength-to-stiffness formulation are applied. In this analysis, the incremental load is carried only by elastic regions. The ratio of local stress increment to macroscopic stress increment $\Delta\sigma / \Delta\Sigma$ is therefore zero for plastic regions. The ratio increases for the remaining elastic regions as yielding progresses. The local stress increment is approximated as being directly proportional to the applied macroscopic stress increment and inversely proportional to the elastic volume fraction raised to an empirical power $n$
\begin{equation}\label{eqn:local_stress_incr}
\Delta \sigma \varpropto \left( \frac{v}{v^e} \right)^n \Delta \Sigma
\end{equation}
where $v^e$ and $v$ are the elastic and total volumes, respectively. The volumetric scaling is an empirical factor that models the increase in local load increment, relative to the macroscopic load increment, for elastic regions that occurs as yielding propagates through an aggregate. The resolved shear stress is a projection of the local stress and therefore exhibits the same scaling
\begin{equation}\label{eqn:rss_incr}
\Delta \tau^* \varpropto \left( \frac{v}{v^e} \right)^n \Delta \Sigma
\end{equation}
Rearranging terms yields
\begin{equation}\label{eqn:drss_dsigbar_scaling}
\frac{\Delta \tau^*}{\Delta \Sigma} \varpropto \left( \frac{v}{v^e} \right)^n
\end{equation}
Introducing the constant of proportionality $\left( d \tau^* / d \Sigma \right)_0$ yields the equality
\begin{equation}\label{eqn:drss_dsigbar}
\frac{\Delta \tau^*}{\Delta \Sigma} = \left( \frac{d \tau^*}{d \Sigma} \right)_0 \left( \frac{v}{v^e} \right)^n
\end{equation}
where $\left( d \tau^* / d \Sigma \right)_0$ is the derivative of the resolved shear stress with respect to the macroscopic stress coefficient at zero load. This derivative can be evaluated from one load increment of a finite element simulation in the completely elastic regime.

Equation~\ref{eqn:drss_dsigbar} provides the basis for estimating the macroscopic load at which crystal volumes will yield during loading.   This is accomplished by integrating Equation~\ref{eqn:drss_dsigbar} numerically over the loading history. The result is the macroscopic stress at which a region yields, corresponding to $\tau^* = \tau_{cr}^*$ in the rate-independent limit
\begin{equation}\label{eqn:rss_integrate}
\tau^* = \sum_{i=1}^N \frac{\Delta \tau^*}{\Delta \Sigma} \Delta\Sigma_i
\end{equation}
When evaluating the macroscopic stress at which elements in a finite element mesh yield, the elements are first binned according to strength-to-stiffness ratio. Binning reduces numerical errors introduced by summing over many small stress increments. Without binning, the number of increments in the summation would be equal to the number of elements in the mesh, since each element has its own unique yield stress. Each bin is associated with a range of macroscopic stresses over which all elements in the bin undergo yield. This range defines the macroscopic stress increment $\Delta\Sigma_i$ for the bin. The average elastic volume is used for the numerical integration.

%% file: Yield_Predict_Application.tex
\begin{figure}[ht]
    \centering
    \subfigure[$n = 0$]
    {\includegraphics[trim = 0in 0in 0.9in 0in, clip, scale = 0.7] {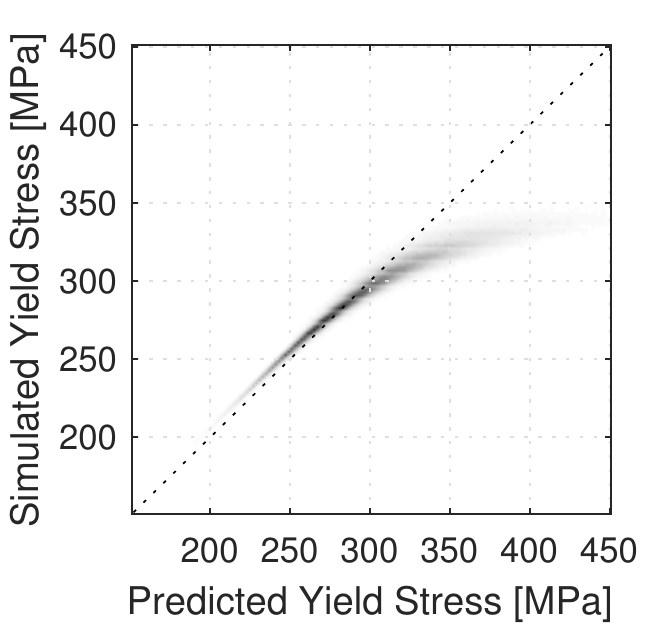}
    \label{fig:VolCorrExp-n000}}
    \subfigure[$n = 2/3$]
    {\includegraphics[trim = 0in 0in 0.9in 0in, clip, scale = 0.7]{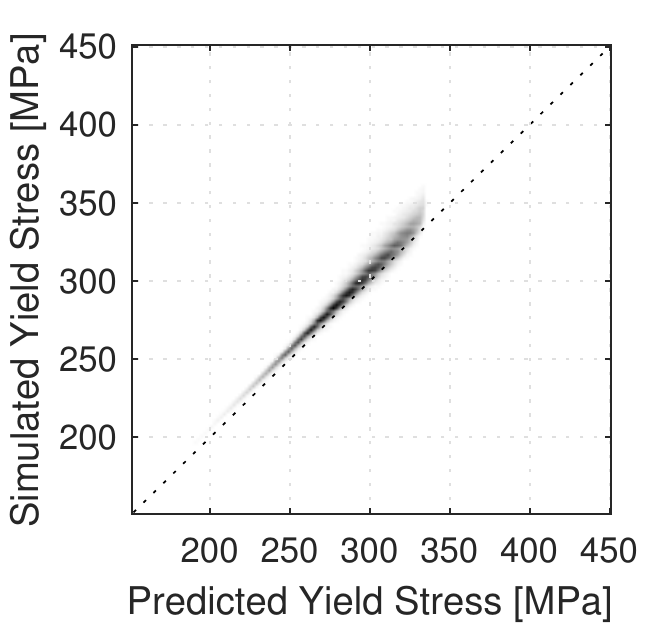}
    \label{fig:VolCorrExp-n067}} 
    \subfigure[$n = 1$]
    {\includegraphics[trim = 0in 0in 0.9in 0in, clip, scale = 0.7]{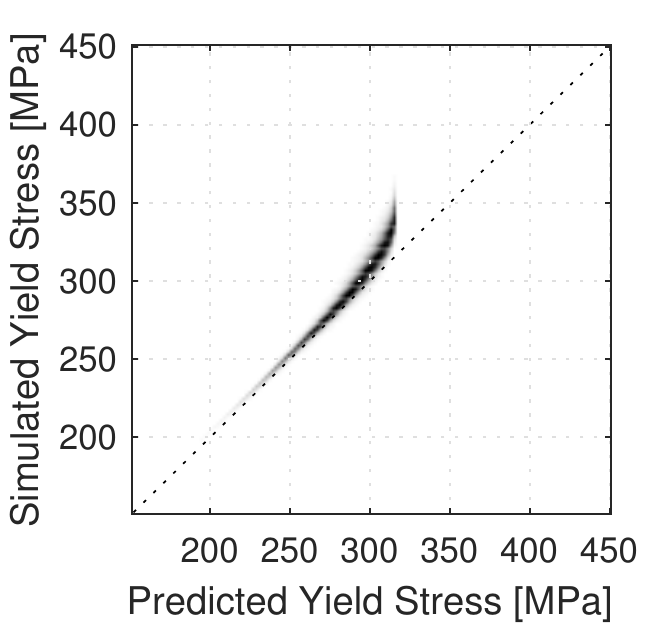}
    \label{fig:VolCorrExp-n100}}
    \caption{Effect of volume correction exponent on predicted elemental yield stress.}\label{fig:VolCorrExp}
\end{figure}

To complete the methodology, the value of empirical power, $n$, acting on the elastic volume fraction in Equations~\ref{eqn:local_stress_incr}-\ref{eqn:rss_integrate} must be specified.
It was determined through the comparison of predicted and simulated macroscopic elemental yield stresses.
Comparisons were made for values of $n$ in the range between zero and unity.
Binned scatter plots showing the relationship between predicted and simulated yield stresses for uniaxial loading are presented in Figure~\ref{fig:VolCorrExp} for three values, $n = 0, 2/3$ and 1. 
The dotted diagonal line in each plot represents the ideal agreement. The first exponent $n=0$ (Figure~\ref{fig:VolCorrExp-n000}) corresponds to no volume correction. Like the strength-to-stiffness plots in Figure~\ref{fig:Y2E} the plot exhibits downward curvature, due to the fact that the local stress increments for elastic regions increase relative to the macroscopic stress increment as yielding progresses. A volume scaling exponent of unity (Figure~\ref{fig:VolCorrExp-n100}), which corresponds to a simple linear scaling,  overcorrects for the volume effect.
Several values in the range between zero and one were also examined,  with the value of $2/3$ providing correlation closest to the ideal (see Figure~\ref{fig:VolCorrExp-n067}). 

\begin{figure}[ht]
    \centering
    \subfigure[$BR = 0.00$]
    {\includegraphics[trim = 0in 0in 0.9in 0in, clip, scale = 0.7]{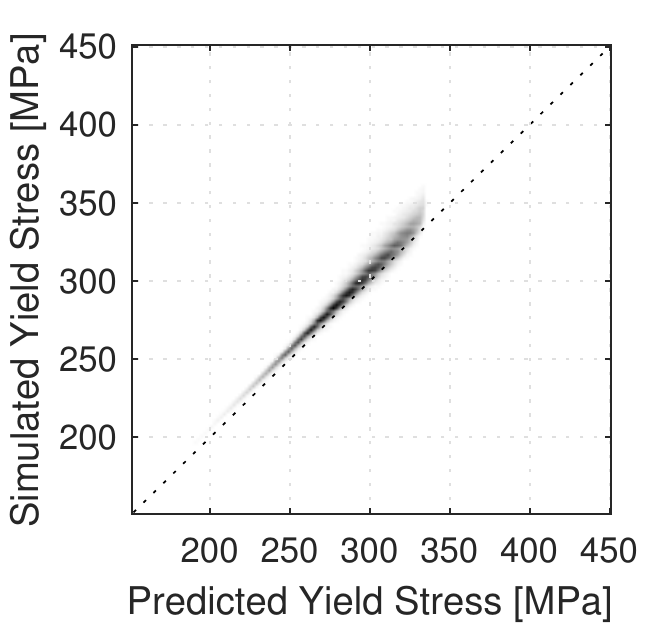}
    \label{fig:ElemYldStress-BR000}}
    \subfigure[$BR = 0.25$]
    {\includegraphics[trim = 0in 0in 0.9in 0in, clip, scale = 0.7]{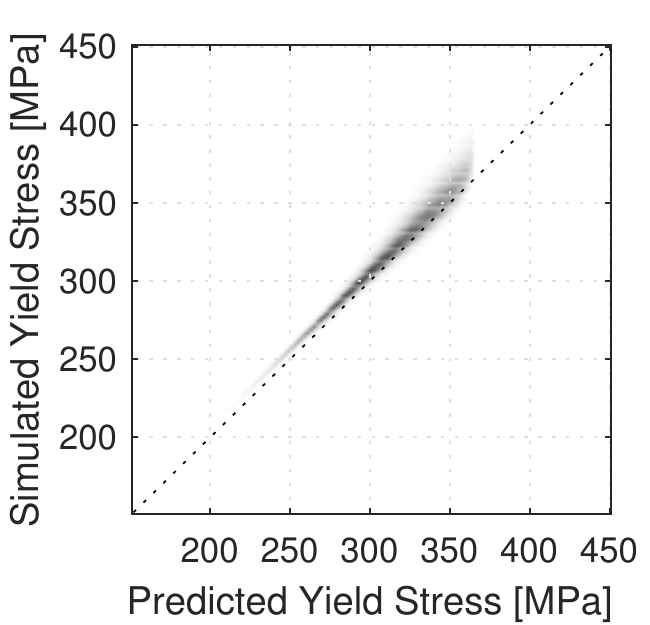}
    \label{fig:ElemYldStress-BR025}}
    \subfigure[$BR = 0.50$]
    {\includegraphics[trim = 0in 0in 0.9in 0in, clip, scale = 0.7]{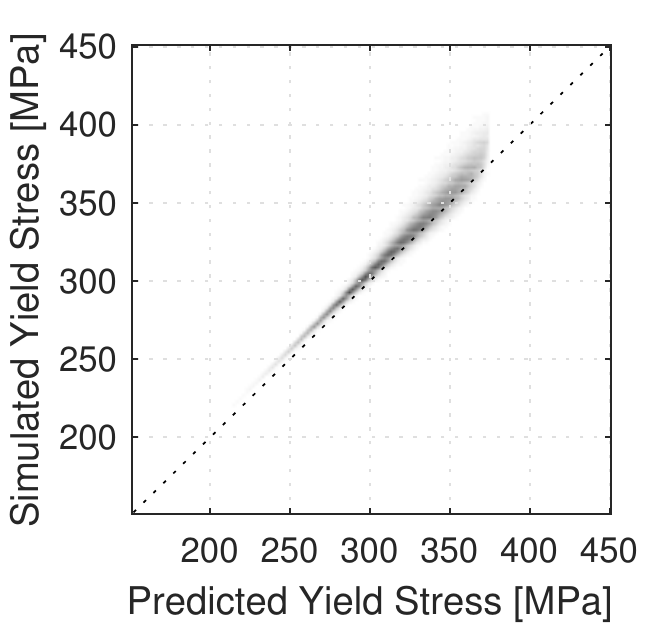}
    \label{fig:ElemYldStress-BR050}}
    \subfigure[$BR = 0.75$]
    {\includegraphics[trim = 0in 0in 0.9in 0in, clip, scale = 0.7]{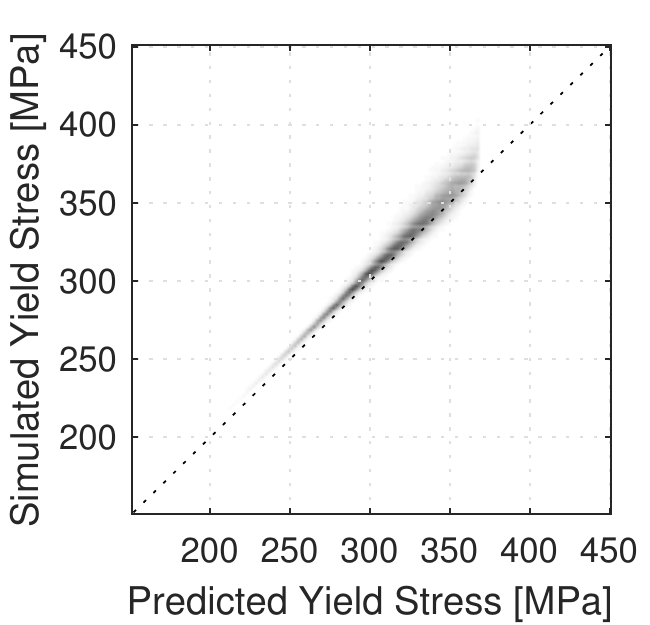}
    \label{fig:ElemYldStress-BR075}}
    \subfigure[$BR = 1.00$]{
    \includegraphics[trim = 0in 0in 0.9in 0in, clip, scale = 0.7]{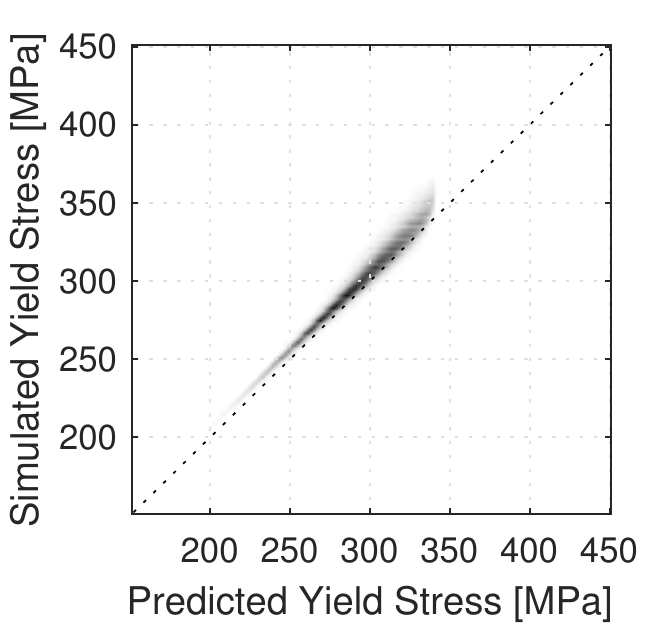}
    \label{fig:ElemYldStress-BR100}}
    \caption{Correlation between the simulated and predicted elemental yield stresses for five levels of stress biaxiality.}
    \label{fig:SimPredElemYldStress}
\end{figure}

The predictive capability of the methodology can now be applied across the range of stress biaxiality.  The predicted yield stress is obtained from Equations~\ref{eqn:local_stress_incr}-\ref{eqn:rss_integrate}
using the stress increments computed for a single step in the elastic regime.   The simulated yield stress values were obtained from the elemental stresses for simulations extending through the elastic-plastic transition.  
Figure~\ref{fig:SimPredElemYldStress} demonstrates the agreement between predicted and simulated macroscopic elemental yield stresses for five levels of stress biaxiality. As with the uniaxial case, there is good agreement between the prediction and simulation, with $R^2$ values greater than 0.93 for all five cases.

\begin{figure}[ht]
    \centering
    \includegraphics[trim = 0in 0in 0in 0in, clip, scale = 0.8]{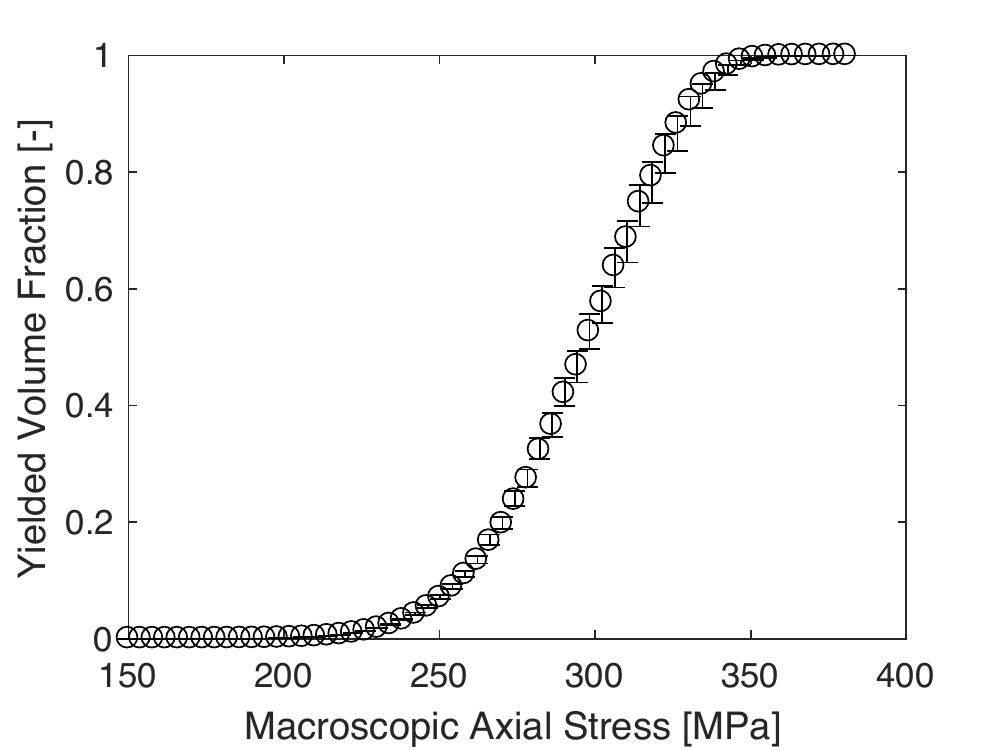}
    \caption{Volume fraction of elements predicted to yield as the load increases and the volume fraction of elements in error.  }
    \label{fig:yieldpredictionerror}
\end{figure}

The predictive power of the new methodology is illustrated with a plot that shows the extent of yielding over a polycrystal with increasing macroscopic stress.  That is, the fraction of the volume that has yielded, and implicitly the complementary fraction that remains elastic, through the elastic-plastic transition.  Again, for the case of uniaxial loading, 
Figure~\ref{fig:yieldpredictionerror} shows the cummulative volume fraction of elements that will yield under increasing load.  Also shown are error bars that indicate  the fraction of elements that are incorrectly predicted to have yielded.  The  volume fraction in error remains small, less than about 0.08, thoughout the loading. This demonstrates the effectiveness of using the directional strength-to-stiffness as the basis for prediction of yielding.


%% file: Conclusions.tex
A generalized metric for the  directional strength-to-stiffness has been proposed for multiaxial stress states.  This metric is an extension of a metric previously developed for uniaxial stress states. The metric is shown to be an effective measure of the mechanical properties of crystalline solids for predicting the volume fraction of crystals yielding under a range of biaxial stress states both for low and high values of elastic anisotropy. It is shown to be more effective than either the Schmid factor or Taylor factor for this purpose as it is incorporates effects of grain interactions that alter the stress during loading.  Using the generalized strength-to-stiffness metric, a methodology is proposed for predicting the progression of yielding in polycrystalline aggregates through the elastic-plastic transition based on stress distributions generated in a single elastic load step. The method is demonstrated for a range of biaxial stress states.   The development presented here is demonstrated only for single-phase alloys, but the approach can be extended to two-phase systems.

%% file: Acknowledgements.tex
Support was provided  by the US Office of Naval Research (ONR) under contract N00014-09-1-0447.

%% file: appendix.tex
The motion of a polycrystal is determined by solving a set of field equations consisting of equations of equilibrium and kinematics, equations for the constitutive behavior, and boundary conditions. 
In this work, we use the finite element code, \fepx, to solve this
set of equations specifically for the mechanical response polycrystalline aggregates.  The summary provided here of the \fepx\, formulation is taken heavily from its documentation~\cite{Dawson14a}.

Equilibrium is enforced by requiring a global weighted residual to vanish:
\begin{equation}
R_u = \int_{\mathcal{B}} \gtnsr{\psi} \cdot \left( \divop {\cauchy}^T + {\vctr{\iota}}\right) \dee \mathcal{B} = 0
\end{equation}
The residual is manipulated in the customary manner (integration by parts and application of the divergence theorem) to obtain the weak form:
\begin{equation}
R_u = 
- \int_{\mathcal{B}} \; \trace  \left( {\dcauchy}^\transp  \,\gradop \vctr{\psi} \right) \dee {\mathcal{B}} 
+  \int_{\mathcal{B}} \pi \, \divop \vctr{\psi} \dee {\mathcal{B}}
+ \int_{\partial {\mathcal{B}}} \vctr{t} \cdot \vctr{\psi} \dee\Gamma +
\int_{{\mathcal{B}}} \vctr{\iota} \cdot \vctr{\psi} \dee{\mathcal{B}}
\end{equation}
Kinematics are introduced through equations that relate the velocity field to the deformation rate and spin via the velocity gradient, as outlined in Section~\ref{sec:model_methods}.  The velocity field is represented with the finite element interpolation functions presented later in the appendix.

The constitutive relations enter to relate the stress to the motion, in particular the deformation rate and spin.  
 Equations~\ref{eq:kinematic_decomp_meandefrate} to \ref{eq:resolved_shear_stress} are now merged into a single equation that relates the Cauchy stress to the
 total deformation rate.   First,  the spatial time-rate change of the elastic strain is approximated with a finite difference expression:
\begin{equation}
\matlatepsdot = \deeteei \biggl( \matlateps - \matlatepsold \biggr)
\label{eq:euler_approx_strain_rate}
\end{equation}
where $\matlateps$ is the elastic strain at the end of the time step and $\matlatepsold$ is the elastic strain at the beginning of the time step.   The difference approximation is employed in an implicit algorithm, wherein the equations are solved at the time corresponding to the end of the time step.  This time corresponds to the current configuration.  
Writing the time rate change of the strain in terms of strains at two times facilitates substitution of Hooke's law -- namely at the end of the time step.   The elastic strain at the beginning of the time step is known from the solution for the preceding time step.
For the volumetric part of the motion this gives:
\begin{equation}
-\pi  =  \frac{\kappa\Delta t }{ \beta } \trace\matdefrate +  \frac{\kappa} {\beta} \trace\matlatepsold
\label{eq:discret_volumetric_ep-law}
\end{equation}
For the deviatoric part, we obtain:
\begin{equation}
\matddefrate = \deeteei \matdlateps + \matlatdefrate + \matpxspinhat \matdlateps - \deeteei \matdlatepsold
\label{eq:discret_deviatoric_ep-law}
\end{equation}
where $\matpxspinhat$ is the matrix form of $\pxspinhat$:
\begin{equation}
\matpxspinhat = 
\left[
\begin{array}{c c c c c}
0 &  0 & -2{\hat w_{12}^p} & -{\hat w_{13}^p} &  {\hat w_{23}^p} \\ 
0 & 0 & 0 & {\sqrt{3}} {\hat w_{13}^p} & {\sqrt{3}} {\hat w_{23}^p}  \\ 
2{\hat w_{12}^p}& 0 & 0 & -{\hat w_{23}^p}  &  -{\hat w_{13}^p}  \\ 
{\hat w_{13}^p}  & -{\sqrt{3}}{\hat w_{13}^p}  &  {\hat w_{23}^p} & 0  &  -{\hat W_{12}^p} \\  
-{\hat w_{23}^p}  & -{\sqrt{3}}{\hat w_{23}^p}  &  {\hat w_{13}^p} & {\hat w_{12}^p}   &  0 
\end{array}
\right] 
\end{equation}
The equations for plastic slip define the plastic deformation rate in terms of the Kirchhoff stress:
\begin{equation}
\matlatdefrate = \matxplasticity \matdkirch
\end{equation}
\begin{equation}
\matxplasticity
= 
\sumss \left( \frac{f(\rss, g)}{\rss} \right) 
\matsymschmid  \matsymschmid^\transp
\end{equation}
When combined with Hooke's law for the elastic response we obtain:
are substituted to render an equation that gives the deviatoric Cauchy stress in terms of the total deviatoric deformation rate and 
a matrix, $\mathhh$, that accounts for the spin and the elastic strain at the beginning of the time step: 
\begin{equation}
\matdcauchy  =  \matxep \bigg( \matddefrate- \mathhh \bigg)
\label{eq:discrete_devCauchy}
\end{equation}
where:
\begin{equation}
\matxep^{\invrs} = \frac{\beta}{\Delta t} {\matxdelasticity}^{\invrs} +\beta \matxplasticity 
\label{eq:discrete_ep_stiffness}
\end{equation}
\begin{equation}
\mathhh = \matpxspinhat  \matdlateps - \deeteei \matdlatepsold
\label{eq:discrete_spin_correction}
\end{equation}
Equations~\ref{eq:discrete_devCauchy}, \ref{eq:discrete_ep_stiffness} and \ref{eq:discrete_spin_correction} are substituted in the weak form of equilibrium to write the stress in terms of the deformation rate.

\fepx\, employs a standard isoparametric mapping framework for discretizing the problem domain and for representing the solution variables.  
The mapping of the coordinates of points provided by the elemental interpolation functions, $\matcapN$, and the coordinates of the nodal points, $\left\{ X \right\}$:
\begin{equation}
\left\{ x \right\}  = \matcapN  \left\{ X \right\}
\label{eq:coord_mapping}
\end{equation}
where $(\xi, \eta, \zeta)$ are local coordinates within an element.  
The same mapping functions are used for the solution (trial) functions which, together with the nodal point values of the
velocity, $\matvelnp$, specify the velocity field over the elemental domains:
\begin{equation}
\matvel  = \matcapN  \matvelnp
\label{eq:trial_functions}
\end{equation}
The deformation rate is computed from the spatial derivatives (derivatives with respect to $\vctr{x}$) of the mapping functions and the nodal velocities as:
\begin{equation}
\matdefrate = \matcapB \matvelnp
\label{eq:trial_function_derivatives}
\end{equation}
$\matcapB$ is computed using the derivatives of $\matcapN$ with respect to local coordinates, $(\xi, \eta, \zeta)$, together with the Jacobian of the mapping specified by Equation~\ref{eq:coord_mapping}, following standard finite element procedures for isoparametric elements.

\begin{figure}[ht]
\begin{center}
\includegraphics*[width=10cm]{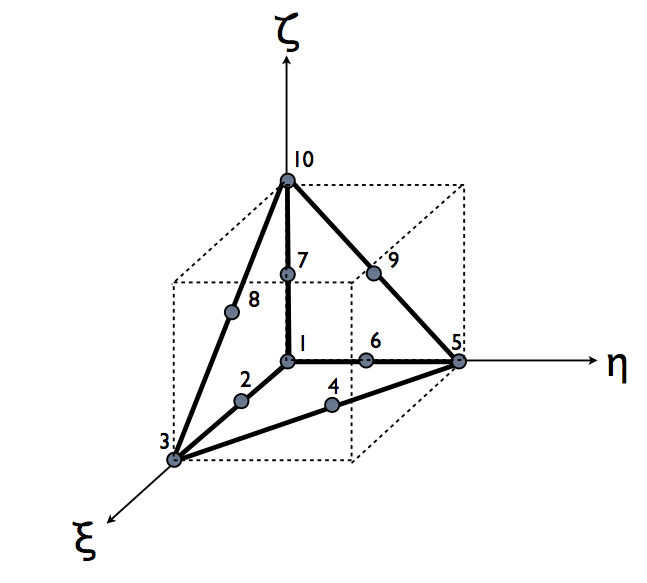}
\caption{10-node tetrahedral element with quadratic interpolation of the velocity, shown in the parent configuration and bounded by a unit cube.}
\label{fig:tet_element}
\end{center}
\end{figure}

\fepx\, relies principally on a 10-node, tetrahedral, serendipity element, as shown in Figure~\ref{fig:tet_element}.  This $C^0$ element provides pure  quadratic interpolation of the velocity field.  
\fepx\, employs a Galerkin methodology for constructing a weighted residual.  The weight functions therefore use the same interpolation functions as used for the coordinate map and the trial functions:
\begin{equation}
\Big\{ \psi \Big\} =  \matcapN  \Big\{ {\it \Psi } \Big\}
\label{eq:weight_functions}
\end{equation}

Introduction of the trial and weight functions gives a  residual vector for the discretized weak form for each element:  
\begin{equation}
\matresiduale
=
\bigg[ \matstiffd + \matstiffv \bigg] \matvelnp
-
\Big\{ \mathsf{f}^{\it ele}_a \Big\}
-
\Big\{ \mathsf{f}^{\it ele}_d \Big\}
-
\Big\{ \mathsf{f}^{\it ele}_v \Big\}
\end{equation}
where
\begin{equation}
\matstiffd 
= 
\int_{\mathcal{B}} \matcapB^\transp  \matcapX^\transp \matxep \matcapX \matcapB \dee {{\mathcal{B}}}
\label{eq:k_d}
\end{equation}
\begin{equation}
\matstiffv = 
\int_{\mathcal{B}} \betaoverkappa \matcapB^\transp \matcapX^\transp \matdelta \matdelta^\transp \matcapX \matcapB \dee{\mathcal{B}}
\end{equation}
\begin{equation}
\Big\{ \mathsf{f}^{\it ele}_a \Big\} 
= 
\int_{\partial{\mathcal{B}}}\matcapN^\transp \matbodyforce \dee {\mathcal{B}}
\end{equation}
\begin{equation}
\Big\{ \mathsf{f}^{\it ele}_v \Big\} 
= \int_{\mathcal{B}} \matcapB^\transp \matcapX^\transp \betaoverkappa \matdelta^\transp \matlatepsold \dee{\mathcal{B}} 
\end{equation}
\begin{equation}
\Big\{ \mathsf{f}^{\it ele}_d \Big\} 
=  \int_{\mathcal{B}} \matcapB^\transp \matcapX^\transp \matxep \mathhh \dee{\mathcal{B}} 
\label{eq:f_d}
\end{equation}
The $\matdelta$ and $\matcapX$ matrices facilitate performing trace and inner product operations.  For details, see \cite{Dawson14a}.
The integrals appearing in Equations~\ref{eq:k_d}-\ref{eq:f_d}  are evaluated by numerical quadrature.

An implicit time integration scheme is used in \fepx\, to advance the solutions for the motion and state over a specified history of loading.  An estimated material state at the end of each time increment is used to evaluate the constitutive equations. Implicit integration ensures stability. The solution algorithm for each time increment begins with initializing an initial guess of the velocity field. The deformed geometry at the end of the time increment is then estimated, based on the velocity field. The velocity gradient is computed and used to solve for the crystal state ($\mathrm{tr} (\tensSym{\epsilon}^e)$, $\tensSym{\epsilon}^{e\prime}$, $\tensSym{R}^*$, and $g^\alpha$) at each quadrature point. Constitutive matrices for the equilibrium equation are computed using the updated material state and used to solve for an updated velocity field. Iteration continues on the geometry, crystal state, and velocity field until the velocity solution is converged. The solution then advances to the next time increment.

%% file: main.bbl
\begin{thebibliography}{1}

\bibitem{Wong10a}
Su~Leen Wong and Paul~R. Dawson.
\newblock Influence of directional strength-to-stiffness on the elastic-plastic
  transition of fcc polycrystals under uniaxial tensile loading.
\newblock {\em Acta Mater.}, 58(5):1658--1678, Mar 2010.

\bibitem{Dawson14a}
P.~R. Dawson and D.~E. Boyce.
\newblock \protect{FEpX} -- \protect{F}inite \protect{E}lement
  \protect{P}olycrystals: Theory, finite element formulation, numerical
  implementation and illustrative examples.
\newblock {\tt arXiv:1504.03296 [cond-mat.mtrl-sci]}, 2015.

\bibitem{Marin98a}
E.~B. Marin and P.~R. Dawson.
\newblock {On modelling the elasto-viscoplastic response of metals using
  polycrystal plasticity}.
\newblock {\em {Comput. Methods Appl. Mech. Engrg.}}, {165}({1-4}):{1--21},
  {Nov} {1998}.

\bibitem{Marin98b}
E.~B. Marin and P.~R. Dawson.
\newblock {Elastoplastic finite element analyses of metal deformations using
  polycrystal constitutive models}.
\newblock {\em {Comput. Methods Appl. Mech. Engrg.}}, {165}({1-4}):{23--41},
  {Nov} {1998}.

\bibitem{SS1}
{Allegheny Ludlum Corporation}.
\newblock Stainless steel al-6xn alloy.
\newblock Technical Data Blue Sheet, no date.

\bibitem{quey_large-scale_2011}
R.~Quey, P.R. Dawson, and F.~Barbe.
\newblock Large-scale \protect{3D} random polycrystals for the finite element
  method: {G}eneration, meshing and remeshing.
\newblock {\em Computer Methods in Applied Mechanics and Engineering},
  200(17--20):1729--1745, 2011.

\bibitem{Marin08a}
T.~Marin, P.~R. Dawson, M.~A. Gharghouri, and R.~B. Rogge.
\newblock Diffraction measurements of elastic strains in stainless steel
  subjected to in situ biaxial loading.
\newblock {\em Acta Mater.}, 56(16):4183--4199, Sep 2008.

\bibitem{Marin12a}
T.~Marin, P.~R. Dawson, and M.~A. Gharghouri.
\newblock Orientation dependence of stress distributions in polycrystals
  deforming elastoplastically under biaxial loadings.
\newblock {\em J. Mech. Phys. Solids}, 60(5):921--944, May 2012.

\bibitem{Ledbetter_1984}
H.~M. Ledbetter.
\newblock Monocrystal-polycrystal elastic constants of a stainless steel.
\newblock {\em Physica Status Solidi (A)}, 85(1):89--96, 1984.

\end{thebibliography}
